\documentclass[a4paper,fleqn,usenatbib]{mnras}

\usepackage{txfonts}
\usepackage[T1]{fontenc}
\usepackage{ae,aecompl}

\usepackage{graphicx}	% Including figure files
\usepackage{amssymb}	% Extra maths symbols

\title[The 1.4\,GHz radio luminosity function]{GAMA/WiggleZ: The 1.4\,GHz radio luminosity functions of high- and
  low-excitation radio galaxies and their redshift evolution to z=0.75}

\author[Michael.~ B.~ Pracy et al.]{
\parbox[t]{\textwidth}{ Michael B.~Pracy$^{1}$\thanks{E-mail: mpracy@gmail.com},
John  H.~Y.~Ching$^{1}$,
Elaine M.~Sadler$^{1}$, 
Scott M.~Croom$^{1}$,
I. K. Baldry$^2$,
Joss Bland-Hawthorn$^{1}$,
S. Brough$^3$,
M. J. I. Brown$^4$, 
Warrick J. Couch$^3$,
Tamara M. Davis$^5$,
Michael J. Drinkwater$^5$,
 A. M. Hopkins$^3$,
 M. J.  Jarvis$^{6,7}$,
 Ben Jelliffe$^1$,
 Russell J. Jurek$^{8}$, 
 J. Loveday$^9$,
 K. A. Pimbblet$^{10,11}$,
 M. Prescott$^7$,
 Emily Wisnioski$^{12}$,
 David Woods$^{13}$},
\\
\vspace*{6pt}\\
$^1$Sydney Institute for Astronomy, School of Physics, University of Sydney, NSW 2006, Australia\\
$^2$Astrophysics Research Institute, Liverpool John Moores University, IC2, Liverpool Science Park, 146 Brownlow Hill, Liverpool, L3 5RF\\
$^3$Australian Astronomical Observatory, PO Box 915, North Ryde, NSW 1670, Australia\\
$^4$School of Physics and Astronomy, Monash University, Clayton, Victoria 3800, Australia\\
$^5$School of Mathematics and Physics, University of Queensland, Brisbane, QLD 4072, Australia\\
$^6$Oxford Astrophysics, Department of Physics, Keble Road, Oxford, OX1 3RH, UK \\
$^7$Astrophysics Group, Department of Physics, University of the Western Cape, Bellville 7535, South Africa\\
$^8$Australia Telescope National Facility, CSIRO, Epping, NSW 1710, Australia\\
$^9$Astronomy Centre, University of Sussex, Falmer, Brighton BN1 9QH\\
$^{10}$E.A.Milne Centre for Astrophysics \& Department of Physics and Mathematics, University of Hull, Cottingham Road, Kingston-upon-Hull,\\ HU6 7RX, UK\\
$^{11}$School of Physics, Monash University, Clayton, VIC 3800, Australia\\
$^{12}$Max Planck Institut fur extraterrestrische Physik, Giessenbachstraβe, D-85748 Garching, Germany\\
$^{13}$Department of Physics and Astronomy, University of British Columbia, 6224 Agricultural Road, Vancouver, BC V6T 1Z1, Canada
}

\date{Accepted XXX. Received YYY; in original form ZZZ}

\pubyear{2015}

\begin{document}
\label{firstpage}
\pagerange{\pageref{firstpage}--\pageref{lastpage}}
\maketitle

\begin{abstract}
We present radio Active Galactic Nuclei (AGN) luminosity functions over the redshift range $0.005 < z < 0.75$. The sample from which the luminosity functions 
are constructed is an optical spectroscopic survey  of radio galaxies, identified from  matched Faint Images of the Radio Sky at Twenty-cm survey (FIRST) sources and Sloan Digital Sky Survey (SDSS) images.The radio AGN are separated  into Low Excitation Radio Galaxies (LERGs) and High Excitation Radio Galaxies (HERGs) using the optical spectra. 
We derive radio luminosity functions for LERGs and HERGs separately in the three redshift bins ($0.005 < z < 0.3$, $0.3 < z < 0.5$  and $0.5 < z <0.75$). The radio luminosity functions
can be well described by a double power-law. Assuming this double power-law shape the LERG population displays little or no  evolution over this
redshift range evolving as $\sim (1+z)^{0.06^{+0.17}_{-0.18}}$ assuming pure density evolution or $\sim (1+z)^{0.46^{+0.22}_{-0.24}}$ assuming pure luminosity evolution.
In contrast, the HERG population evolves more rapidly, best fitted by $\sim (1+z)^{2.93^{+0.46}_{-0.47}}$ assuming a double power-law shape and pure density
evolution. If a pure luminosity model is assumed the best fitting HERG evolution is parameterised by $\sim (1+z)^{7.41^{+0.79}_{-1.33}}$. The characteristic break 
in the radio luminosity function occurs at a significantly higher power ($\gtrsim 1$\,dex) for the HERG population in comparison to the LERGs. This is 
consistent with the two populations representing fundamentally different accretion modes. 
\end{abstract}

\begin{keywords}
galaxies: evolution, accretion -- galaxies: active -- radio continuum: galaxies
\end{keywords}

%%%%%%%%%%%%%%%%%%%%%%%%%%%%%%%%%%%%%%%%%%%%%%%%%%

%%%%%%%%%%%%%%%%% BODY OF PAPER %%%%%%%%%%%%%%%%%%
\section{Introduction}
The  evolution of galaxies and the supermassive
black holes at their centres appear to be closely connected.  This connection
is evident in the correlation between black hole mass and the
stellar bulge mass \citep[][]{magorrian98} and black hole mass and stellar velocity dispersion \citep[e.g.][]{gebhardt00}.
One way this coupling may occur is via  feedback processes from the
Active Galactic Nuclei (AGN) depositing energy, released by the accretion of
matter onto the black hole, into their host galaxy and its
surrounding environment.  
For example, radio jets from AGN are commonly invoked as a feedback mechanism to inhibit
gas cooling and suppress star-formation in massive galaxies and
clusters of galaxies \citep[e.g.][]{binney95,fabian02,best06,croton06,bower06,mcnamara07}. Such a feedback cycle can
simultaneously solve the cooling flow problem; the sharp bright-end turn down in the optical galaxy luminosity
function and the fact that  the most massive bulge dominated galaxies
contain the oldest stellar populations \citep{croton06}.\\

The properties of the AGN population indicate that there are two
fundamentally different accretion modes operating
\citep{hardcastle07,best12}.  In the classical `cold mode', material
is accreted onto the supermassive black hole via a small, geometrically thin, optically
luminous accretion disc. This disc is the source of ionising photons
producing both broad and narrow line emission in the optical spectrum
of AGN  and x-ray emission via the inverse-Compton process. In the
unified AGN model \citep{antonucci93}, the broad line region is obscured by a dusty torus
when viewed from certain orientations resulting in the type I (not
obscured) and type II (obscured) AGN classifications.  As a result of the emission lines in their optical spectra, these `cold-mode' AGN
 are also referred to as `quasar mode' or `high-excitation' AGN.
However, there  is a large population of AGN observable by their radio emission
but without the bright high-ionisation emission lines in their optical spectra
\citep[e.g.][]{hine79,laing94,jackson97,best12}. These display no evidence for the presence of an accretion
disc or dusty torus \citep[e.g.][]{Chiaberge02,Whysong04} and are likely powered by
radiatively inefficient accretion possibly from a hot gas halo \citep[e.g.][]{hardcastle07,best12}.
These `hot-mode'  AGN (also referred to as `radio mode' or `low-excitation'
AGN) are generally more massive, have higher mass-to-light
ratios, are redder, and are of earlier morphological type than strong-lined AGN 
\citep[e.g.][]{kauffmann03, kauffmann08, best12}.
The `cold-mode' AGN display a higher rate of interactions and
peculiarities in their morphology  \citep[e.g.][]{smith89}. 

Both AGN modes will inject energy into their environment and have the
potential to influence star-formation and galaxy evolution. In the
`hot mode' the radio jets heat the surrounding hot gas
atmosphere. This is the same gas that fuels the AGN allowing a self
regulating feedback cycle with a balance established between heating and cooling
\citep{croton06,best06}. In the `cold mode' AGN are known to drive
winds in their host galaxy  \citep[e.g.][]{ganguly08,harrison14,mcelroy15}  and
have been suggested as mechanisms to both enhance \citep{silk10} and
suppress \citep{shabala11,page12,davis12} star formation. 

In the local Universe the number of radio AGN with low-excitation
optical spectra (Low Excitation Radio Galaxies, hereafter LERGs) out
number the High-Excitation Radio Galaxies (hereafter, HERGs) at all
but the highest 
radio luminosities \citep{best12}.  The change in the dominant
population occurs at  $L_{\rm 1.4\,GHz} \sim 10^{26}{\rm W~Hz}^{-1}$.
The cosmic evolution of the space density of  radio AGN is sensitive
to radio luminosity. It is well established that the density of the most powerful radio AGN 
increases rapidly with increasing redshift out to $z\sim 2$
\citep{longair66,doroshkevich70,willott01,sadler07} -- 
increasing by a factor of $\sim 1000$.  At higher redshift the space density flattens and then decreases. 
The decrease occurs at higher redshift for more luminous sources \citep[e.g.][]{peacock85,dunlop90,cirasuolo06}. 
This fast evolving high luminosity population is mostly associated with Fanaroff-Riley
II radio sources \citep{jackson99} and objects with strong
optical emission lines \citep{willott01,best12} i.e. HERGs. Conversely, the low luminosity radio
sources are more commonly associated with Fanaroff-Riley
I sources and objects which lack strong optical emission lines
i.e. LERGs. The space density of these low luminosity radio galaxies shows
little (a factor $\sim 2$) or no redshift evolution 
out to z$\sim 1$
\citep[e.g.][]{clewley04,sadler07,donoso09,smolcic09,mcalpine13}. 

\citet{best12} constructed the first radio galaxy luminosity function
with the LERG and HERG contributions separated. They found, that while the LERGs outnumber
the HERGs below $L_{\rm 1.4\,GHz} \sim 10^{26}{\rm W~Hz}^{-1}$,
examples of both classes are found at all radio luminosities. The HERG
population shows evidence of evolution  over the
redshift range of their sample ($z<0.3$) while there is no evidence
for cosmic evolution of the LERG population. This implies the evolution of the radio
luminosity functions dependence on radio luminosity can, at least in part, be explained by a
changing mix in the populations \citep{best12}. \citet{best14} using a composite of 8 published AGN samples
\citep{wall85,gendre10,rigby11,lacy99,hill03,waddington01,simpson06} constructed a catalogue of 211 radio-loud AGN in the
range $0.5 < z < 1.0$. Using this to compare with local samples \citep{best12,heckman14} they made the first measurement of
the evolution of the radio AGN population separated,
spectroscopically, into  LERGs (jet-mode in their nomenclature) and HERGs (dubbed radiative-mode).  They found
that the space density of the HERGs in their sample increased by around an order of
magnitude to z=1. In contrast, the LERG AGN density decreased with 
redshift at luminosities below $L_{\rm 1.4\,GHz} \sim 10^{26}{\rm W~Hz}^{-1}$ and increased at higher
luminosities.

In this paper we present the radio luminosity function of LERGs and
HERGs and their cosmic evolution to z=0.75, corresponding to the second
half of the history of the Universe.  These luminosity functions are
based on a new sample of over five thousand radio galaxies ($0.005 < z
<0.75$) with confirmed spectroscopic redshifts and spectroscopic
classifications. Throughout this paper we convert from observed to physical units
assuming a $\Omega_{M}=0.3$, $\Omega_{\Lambda}=0.7$ and
$H_{0}=70$\,km\,s$^{-1}$\,Mpc$^{-1}$ cosmology.

\section{Sample construction}
The underlying sample from which we constructed the luminosity
distributions of various radio galaxy populations is the Large Area
Radio Galaxy Evolution Spectroscopic Survey
\citep[LARGESS;][]{ching15}. 
This sample was constructed by matching the Faint Images of
the Radio Sky at Twenty-cm survey \citep[FIRST;][]{becker95} with the
photometric catalogue of the Sloan Digital Sky Survey data release 6
\citep[SDSS;][]{adelman08}. The catalogues were matched to all objects
in the FIRST catalogue ($\lesssim 0.5$\,mJy) and an optical apparent magnitude limit of i=20.5
(SDSS $i$-band model magnitudes). The matching procedure was designed to
include both point and extended radio sources, including those with
multiple components. This matching was performed over $\sim$900 square degrees of
sky in regions where a high completeness of optical
spectroscopy could be obtained. A careful estimation of the matching reliability and completeness was performed by \citet{ching15}
using monte-carlo simulations of randomised catalogs. The overall reliability of the matching is estimated as $\sim 93.5$\,per cent and 
the overall completeness is $\sim$95\,per cent. The incompleteness and inclusion of false matches will somewhat offset in the effects on the 
normalisation of the luminosity functions. We do not do any scaling of the luminosity functions to account for input catalog incompleteness but note
that its effect on the normalisation could not be more than a few per cent. 

A possible systematic could arise if the incompleteness or false matches are biased toward a particular source type or source properties. The 
most likely source of such a bias is the behaviour of the matching algorithm for sources with complex (non-point source) radio morphologies. Such sources
are more common at low redshift and lower luminosity as a result of angular resolution effects and the presence of extended star-forming galaxies. 
There is no evidence for such a bias. Only $\sim$10 per cent of the sample is matched to more than one FIRST component, which is consistent with other
determinations from the literature \citep{ivezic02}. Of  those $\sim$10 per cent the fraction of matches to the random catalogs are almost identical for FIRST sources 
with two, three and four or more  components. 

While the high spatial resolution of the FIRST catalogue makes it preferable for matching to the optical photometry it will also resolve 
out  extended radio emission resulting in lower flux density measurements. Since  angular size decreases with redshift this flux loss
will be more significant at low redshift and could mimic evolution in the radio luminosity function. To avoid this we replace  the flux densities with those  measured from the
lower spatial resolution NRAO VLA Sky Survey  \citep[NVSS;][]{condon98} catalogue. In the case where a single NVSS source 
coincides with more than one FIRST source the NVSS flux is split between objects with the same flux
ratio as measured by FIRST (see \citet{ching15} for details of the matching of FIRST and NVSS sources). 

At the faintest flux densities the NVSS catalogue will suffer from significant incompleteness. To avoid
working with a parent sample that suffers from such incompleteness we
only include objects with flux density greater than  2.8\,mJy in constructing our luminosity function. 
This is the same flux limit adopted by \citet{sadler07} when matching radio
sources to the luminous red galaxies in the 2dF-SDSS LRG and QSO survey
\citep[2SLAQ;][]{croom09}. With this flux-density limit the source
counts are still rising at the faintest flux densities; as demonstrated
in Figure \ref{fig:counts}. This results in a sample of 
10827  matched radio/optical sources. 
\begin{figure}\setcounter{figure}{0}
      \includegraphics[width=6.4cm, angle=90]{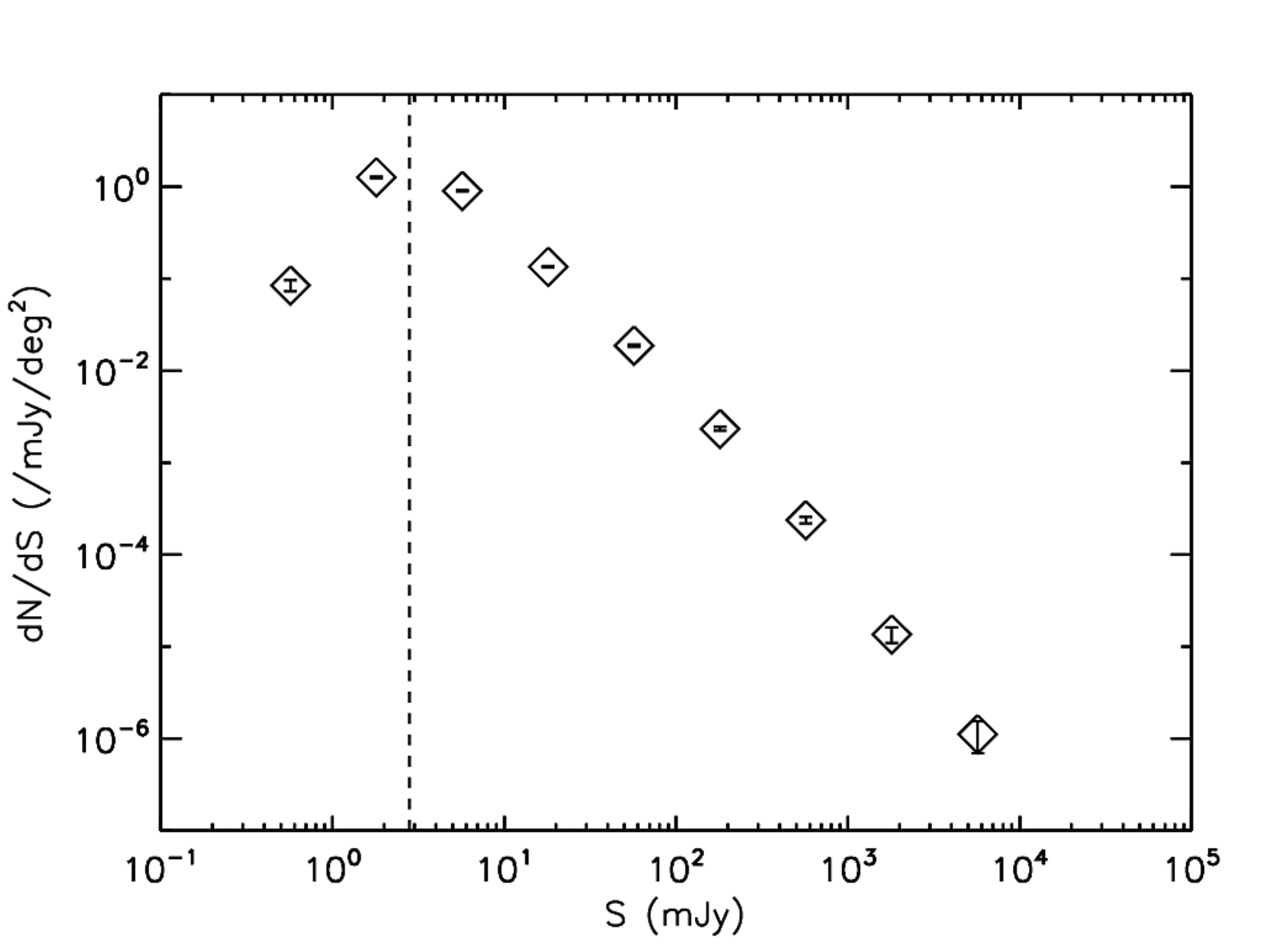}
\caption{\label{fig:counts} Number of objects per interval of 1.4\,GHz
  flux density in the parent sample ({\it diamonds}). The vertical {\it dashed line}
  shows the flux density limit ($S_{1.4GHz} > 2.8$\,mJy) applied in
  this paper. The number counts are still rising at this flux density.}
\end{figure}

The optical spectroscopy for the sample was gathered either from
publicly available archival sources such as the SDSS \citep{adelman08}, 2SLAQ
\citep{cannon06,croom09} and the
2dF QSO Redshift survey \citep[2QZ;][]{croom04} or from a dedicated
`spare-fibre' campaign 
using the 2dF/AAOmega instrument on the Anglo Australian Telescope. This was done by
`piggy-backing' on the WiggleZ Dark Energy Survey  \citep[WiggleZ;][]{drinkwater10} and GAMA
\citep{baldry10,driver11,liske15} large survey programs by using a small number of
fibres on each survey field to target radio galaxies (see \citet{ching15} for details). This has almost
no impact on the large survey efficiency but over  time results in a large number of `spare-fibre' spectra being obtained. 
Of the 10827 radio galaxies satisfying our 1.4\,GHz flux density
criterion, the number with optical spectra is 7088. Of these 6215 are deemed to have  reliable redshift determinations ($\sim$87\,per cent).

\subsection{Optical spectroscopic classification}\label{sec:class}
In addition to providing an accurate measure of the redshift the
optical spectroscopy can be used to determine the physical origin of the
radio emission i.e. star-formation or AGN. The optical spectroscopy
can also be used to separate the AGN into LERGs and HERGs. Full
details of the semi-automatic classification procedure is given in
\citet{ching15}. In brief, FIRST galaxies whose radio emission is
dominated by star-formation are identified using a Baldwin, Philips and Terlevich 
\citep[hereafter BPT;][]{baldwin81} diagram. Only galaxies with radio luminosity
$L_{1.4GHz}\le 10^{24}$\,W\,Hz$^{-1}$ and $z<0.3$ are considered,
since galaxies more powerful than this would require unrealistically
high star-formation rates and so are assumed to have their radio emission generated by an AGN. The BPT diagnostic will not work in the
case where the galaxy has a radio-quiet  optical AGN and radio
emission powered by star-formation \citep{best12,ching15}. These cases
were identified based on a comparison of the inferred star-formation rates
from H$\alpha$ and radio luminosity. The remaining radio galaxies are
classified as AGN and further subdivided using the equivalent width of
the [OIII]$\lambda 5007$ emission line, where galaxies are classified
as HERGs if they have  SNR([OIII]$\lambda 5007) > 3$ and EW([OIII]$\lambda 5007) > 5$\,\AA. 
See \citet{ching15} for a detailed description of the methods used to make the line strength measurements. 
The 5\,\AA\, demarcation is the same as that used by \citet{best12}.

The O[III]$\lambda 5007$ line is redshifted out of the WiggleZ and
SDSS spectra at $z \sim 0.83$ and out of the GAMA spectra at $z \sim 0.76$.
These also correspond, approximately, to the
redshift where our optical k-corrections are expected to be reliable
\citep{blanton07}. In addition, as a result of the magnitude limit of the optical spectroscopic follow up, above z$\sim 0.75$  the spectroscopic sample is
dominated by broad line AGN and there are almost no narrow-line AGN or 
LERGs in the sample (see Figure \ref{fig:zhist}). We therefore impose a redshift limit of $z=0.75$ 
for construction of our luminosity functions. We also apply a lower
redshift limit of $z=0.005$ to remove Galactic objects. With these  cuts
the final number of objects, with reliable redshifts, is 5026.  A summary of the number of spectra from each survey in this final 5026 objects is given in Table \ref{tab:specsource}.
\begin{table*}
\caption{A summary of the number of spectra from each survey in this final 5026 objects which are used in the measurement of the luminosity functions.}
\begin{center}
\begin{tabular}{lcccc}
\hline
                             &           $z<0.3$     &      $0.3\le z< 0.5$     &   $0.5\le z <0.75$      &            Total                         \\ \hline
N(SDSS)              &          1958               &         846                       &           221                   &        3025                              \\
N(WiggleZ main)$^a$ &            26                 &         137                        &        151                            &  314                                          \\
N(WiggleZ radio)$^b$ &            72                 &         354                        &        389                            &   815                                         \\         
N(WiggleZ other)$^c$ &            9                   &         15                          &         40                           &      64                                   \\
N(GAMA)            &            154                 &        298                         &      221                         &   673                                   \\
N(2SLAQ LRGs)   &             0                  &           32                       &             95                   &          127                           \\
N(2SLAQ QSOs)  &              1                 &            1                       &               3                   &             5                            \\
N(2QZ/6QZ)       &               1                &              1                     &               1                  &             3                          \\
Total                  &             2221             &           1684                    &           1121              &                          5026          \\
\hline
 \end{tabular}
\label{tab:specsource}
\end{center}
$^a$WiggleZ main survey targets.\\
$^b$Spare fibre targets for this project.\\
$^c$Target for other spare fibre programs.\\
\end{table*}

\begin{figure*}\setcounter{figure}{1}
    \begin{minipage}\textwidth
      \includegraphics[width=6.4cm,angle=90]{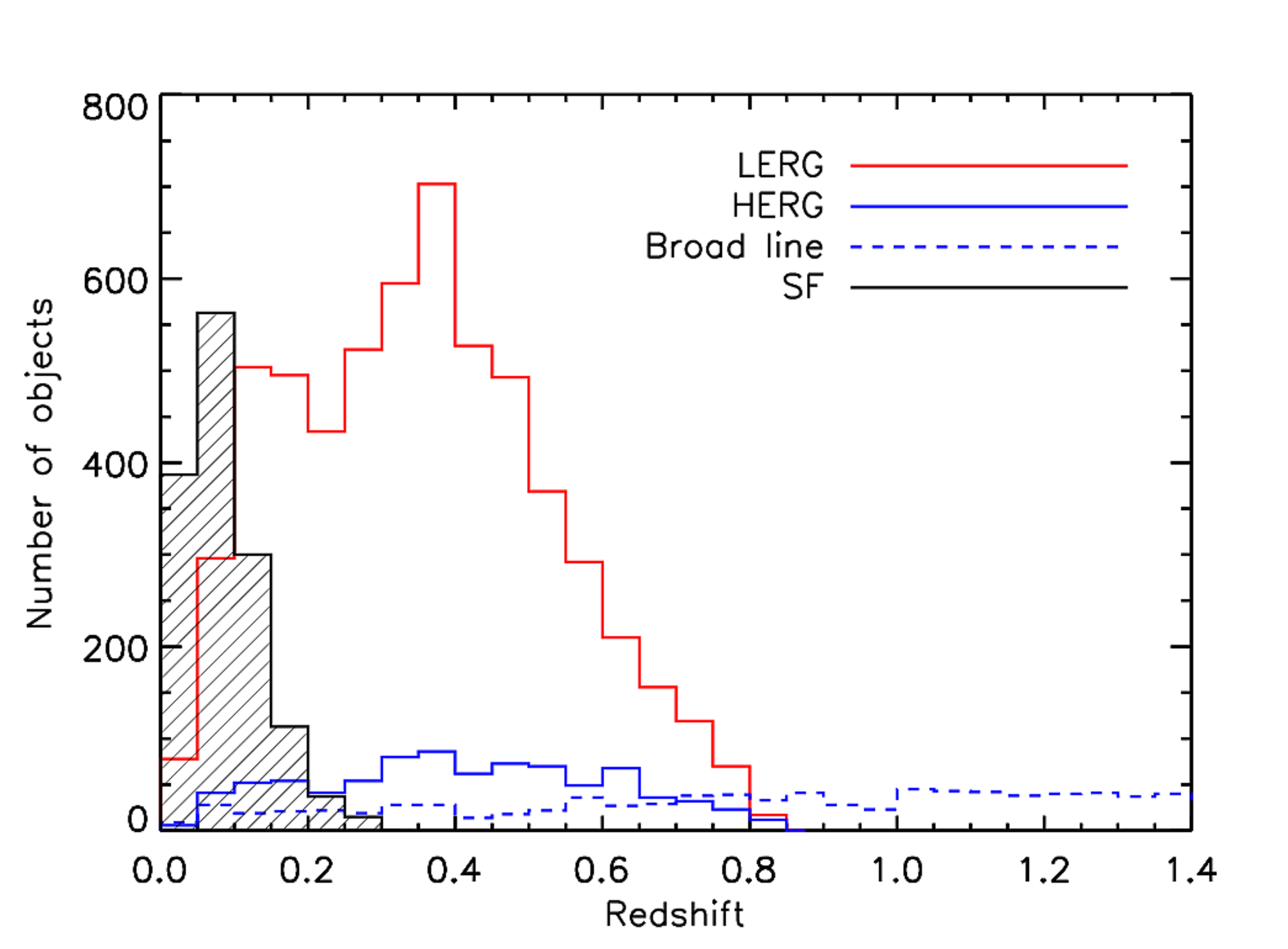}
      \includegraphics[width=6.4cm, angle=90]{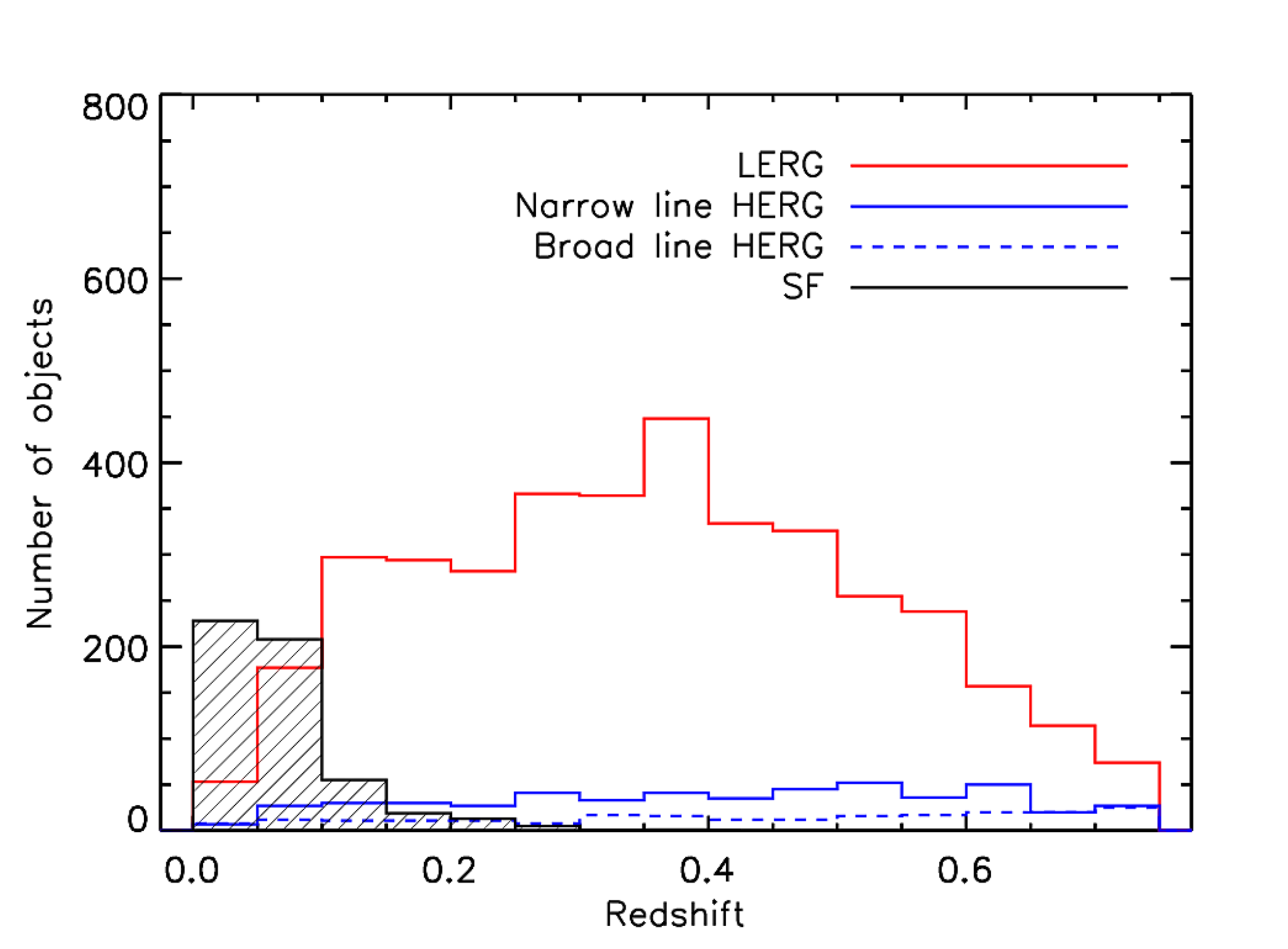}
      \end{minipage}
\caption{\label{fig:zhist} {\it Left panel:} The redshift distribution of all radio galaxies
with reliable spectroscopic redshifts. The
redshift distributions for star-forming galaxies, LERGs and HERGs are
shown separately. Above $z\sim 0.8$ the sample is dominated by broad-line AGN. {\it Right panel:} The n(z) of the sample used to
construct luminosity functions in this paper after selecting objects with $0.005<z<0.75$ and  $S_{1.4GHz} > 2.8$\,mJy.}
\end{figure*}
Not all objects can be classified using the semi-automated
methods of \citet{ching15}. Out of the 5026 objects in our sample 284
($\sim 6.7$\, per cent) were not  classified by \citet{ching15}. A
fraction of the spectra in the \citet{ching15}  sample had been inspected
and classified visually, and in the cases where there is no automated
classification but a visual one, we use this visual
classification. This leaves 229 objects (corresponding to $\sim 4.6$\,
per cent) without a classification. Rather than carry these unclassified objects through the analysis we performed visual classification of these sources. 
For the most part this was straight-forward.
The fraction of objects classified as each type visually are similar to the overall fractions but with a lower frequency of  `star-forming' objects and a higher fraction of HERGs.
The fractions of objects of each type prior to the visual classification are $\sim$75.2, 14.3 and 10.5\, per cent for the LERGs, HERGs and star-forming galaxies, respectively.  The corresponding 
fractions in the visual classification are $\sim$78.6, 20.9 and 0.5\, per cent. 
The increase in the HERG fraction (including quasars) and the decrease in the star-forming fraction is expected since the redshift distribution 
of the unclassified objects is, not surprisingly, peaked at the high redshift end of the sample.

\section{The radio luminosity function}
We first wish to construct the bivariate luminosity function for all galaxies in
the optical-radio matched sample in multiple redshift bins. That is, we
calculate the volume density of galaxies per
interval of radio luminosity per interval of optical luminosity in each
redshift bin. To do this we use the standard $1/V_{\rm max}$ method \citep{schmidt68},
where the number density in each bin is given by:
\begin{equation}
\Phi (M_{\rm i},L_{\rm 1.4GHz}) \Delta M_{\rm i }\Delta \log L_{\rm
  1.4 GHz}= \sum_{\rm gal} {w_{\rm gal} \over V_{\rm max}(\rm gal)}\label{eq:lf}
\end{equation}

The sum on the right-hand side is over all galaxies in the
bin. The weight, $w_{\rm gal}$,  given to each galaxy 
corrects for incompleteness in the spectroscopic follow up. For a
complete survey $w_{\rm gal} = 1$ and for incomplete samples $w_{\rm
  gal}$ is given by the reciprocal of the completeness. The
denominator, $V_{\rm max}({\rm gal})$, is the maximum volume over which
the galaxy could have been observed given the selection limits in both the radio and optical.  We
outline our methods for estimating $w_{\rm gal}$ and   $V_{\rm
  max}({\rm gal})$ below.

\subsection{Estimating the completeness}
The overall spectroscopic completeness is given by the number of
targets for which we obtained a spectrum of sufficient quality to make
a reliable redshift measurement divided by the number of targets in the
parent radio/optical matched sample. That is  6215/10827 or
$\sim$57\,per cent. This is mostly targeting incompleteness; the 
percentage of spectroscopically targeted objects which resulted in a good 
quality redshift classification, i.e. spectroscopic completeness, is $\sim 90$\,percent.
This spectroscopic incompleteness is not random. It will depend sensitively
on the optical magnitude for at least two reasons. Firstly, the
optical spectroscopy of the sample is constructed from spectroscopy
from multiple surveys, which have different limiting
magnitudes. In addition, the spectroscopic completeness within an individual survey
 generally decreases with optical magnitude as the noisier spectra
 obtained for fainter galaxies make it increasingly difficult to
 identify the redshift.  \citet{ching15} used repeat observations to demonstrate that there is little difference in the likelihood
of obtaining a redshift for different spectral classes (with and without emission lines) once optical magnitude is taken into account. 
They did this by comparing the fraction of objects of different spectral classes that required more than one repeat observation
to acquire a good redshift. The idea being that if for a particular source type it is easier to identify a redshift (at give apparent magnitude) than
it should, on average, require less repeat observations. While they did find a slightly higher tendency for spectra without emission lines 
to require a repeat observation than those with emission lines, it was not statistically significant. One important caveat on this technique is 
that it can only be used in cases where a reliable redshift is eventually obtained by one of the observations.

The completeness will also be
 sensitive to colour, again for at least two distinct reasons. Firstly,
 while most of spectra were obtained from surveys that do not apply
 colour selection criteria (e.g. our dedicated follow-up, targets with
 SDSS spectra and the GAMA main survey), a fraction of the spectra
 come from surveys that do (e.g. the WiggleZ main survey, 2SLAQ and 2QZ). 
Another reason completeness will vary with colour, is that colour correlates with other spectral
properties which make a redshift easier to identify. In particular,
galaxies with bluer colour are more likely to have easy to identify
emission lines. Although, as mentioned above, this appears to cause little bias.

There may be other properties which influence the spectroscopic
completeness (e.g  changes in the observable spectral features
in a fixed observing band with redshift) but it is likely that magnitude and colour account for the
dominant dependencies. We therefore choose to construct our
spectroscopic completeness in this plane, specifically we calculate
our completeness (and hence our $w_{\rm gal}$'s) in the $i$-magnitude
versus $g-i$ plane.

We calculate the completeness for each object (or position) in the $i$-magnitude versus $g-i$ plane by averaging 
over an adaptively sized circle in this plane. The size of the circle is calculated by using the distance in the plane 
to the 50th nearest neighbour in the photometric input catalog. Using this adaptive bin size for estimating the completeness 
means in regions of the plane with few objects we still use a reasonable number of objects in estimating the completeness, while
 in high density regions we obtain a finer more local sampling of the completeness.  The completeness in this plane is shown in 
Figure \ref{fig:complete}. Each galaxy is assigned a $w_{\rm gal}$ given by the reciprocal of the completeness. 
Using this nearest neighbour method for estimating the completeness even objects with large weights have had their completeness estimated 
using a reasonable number of objects. For example, for an object with a weight of 10 (completeness of 0.1) the estimate is based on five 
objects with good spectra and 50 objects in the input catalogue. We also checked that our luminosity functions are not largely affected 
by objects with the highest weights. Removing objects with weights greater than 10 makes no significant quantitive difference and
 no qualitative difference to our results. 
\begin{figure*}\setcounter{figure}{2}
      \includegraphics[width=13.8cm, angle=90]{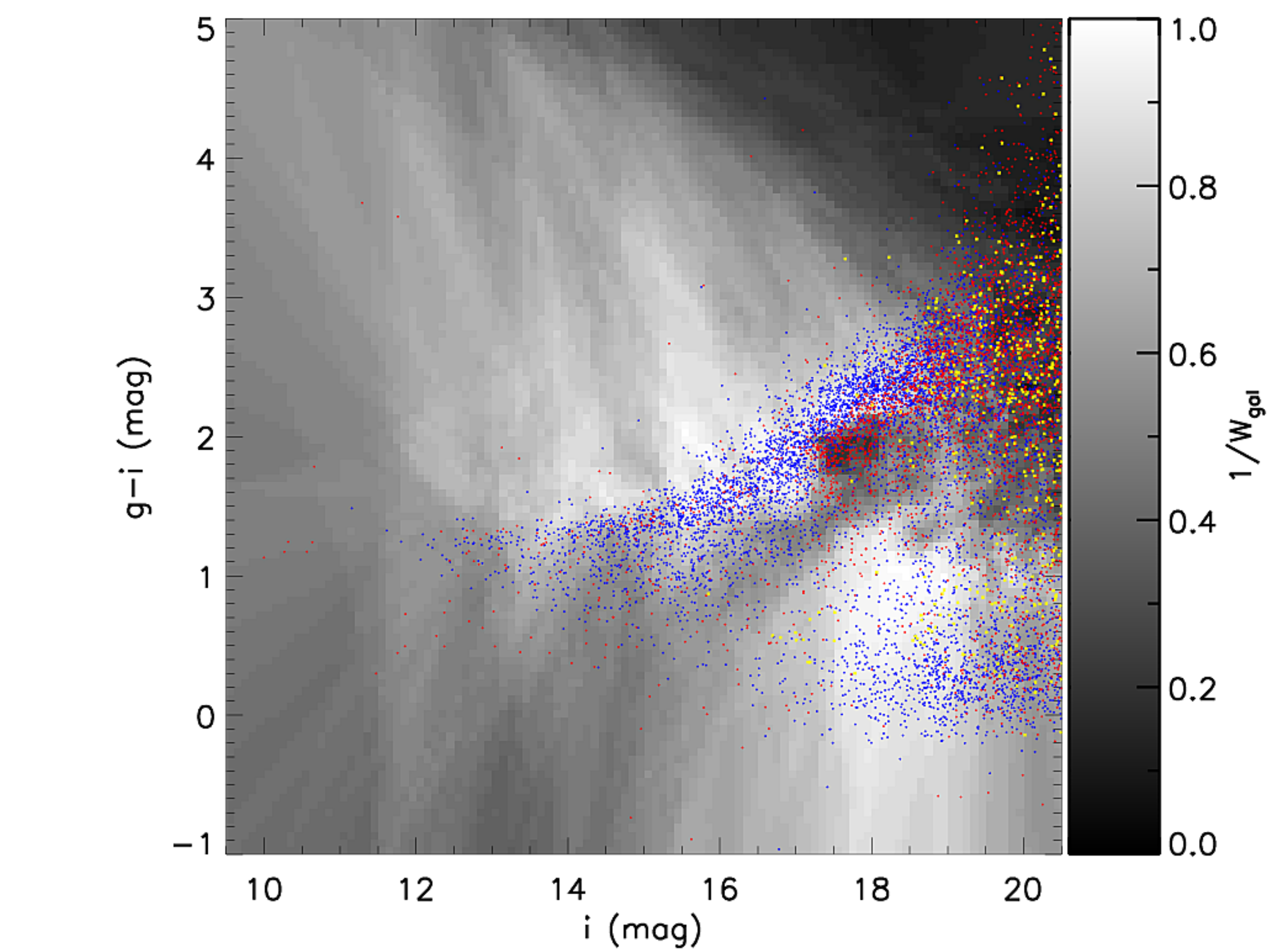}
\caption{\label{fig:complete} The completeness map in the $i$-magnitude versus $g-i$ colour plane,  used to generate the weight ($w_{\rm
    gal}$; shown in sidebar) assigned to each galaxy in constructing the luminosity
  function.  Overlaid {\it blue points} are the photometric sample objects above our 2.8\,mJy flux limit for which a reliable redshift has been obtained.  The overlaid {\it red points} are objects for which no spectroscopic follow observation was made.  The {\it yellow points} represent objects 
which were observed spectroscopically but a reliable redshift could not be determined. 
The structure in the completeness map, such as the low completeness for faint objects with $g-i$ colours of $\sim $1--2\,mag, is 
mostly the result of targeting completeness in the constituent surveys and demonstrates the importance of constructing our completeness estimates in this plane.}
\end{figure*}

\subsection{Estimating $V_{\rm max}$}
To evaluate Equation \ref{eq:lf}, and correctly weight the contribution to the
luminosity function of each galaxy in our sample, we need to estimate
the volume within which each galaxy would satisfy the optical  and  1.4\,GHz flux density selection
limits i.e. $V_{\rm max}$.   This volume will depend on the redshift
of the source, its optical magnitude, 1.4\,GHz flux density, and the
spectral shape in both the optical and radio regimes. To do this, for each
source, we first calculate the absolute $i$-band magnitude including a
k-correction and evolutionary correction:
\begin{equation}
M_{i}=m_{i}-{DM}-K_i(z)-e(z)\label{eq:absmag}
\end{equation}
where ${DM}$ is the distance modulus,  $K_i(z)$ is the k-correction and $e(z)$ is the evolutionary correction (e-correction). The
k-correction is calculated using version v4\_2 of the   {\sc kcorrect}
package \citep{blanton07} using the 5 band (u,g,r,i,z) SDSS photometry. 
The e-correction accounts for the fading of stellar populations with time. For the LERGs we
use the e-corrections for early type galaxies from \citet{poggianti97}. The HERGs are generally bluer, 
more likely to be interacting and star-forming (or recently star-forming) systems which mitigates the effect 
on the overall optical luminosity of the ageing of the underlying older stellar population. Given this we
set $e(z) = 0$ for the HERGs and discuss the effects this assumption has on the evolution of the luminosity function in Section \ref{sec:discussion}.
 It is worth noting the magnitude of the e-correction is significant: increasing from $\sim$0.1\,mag (i-band) at z=0.1 to $\sim$0.8\,mag at z=0.75.

We then calculate the radio luminosity, also including a k-correction:
\begin{equation}
L_{\rm 1.4GHz}=4 \pi d_{\rm L}^{2} {1 \over (1+z)^{(1+\alpha) }}\label{eq:radiopower}
S_{1.4{\rm GHz} }
\end{equation}
here $d_{L}$ is the luminosity distance and $\alpha$ is the spectral index,
defined as $S_{\nu}=\nu^{\alpha}$. We assume the canonical $\alpha =
-0.7$ \citep{sadler02,condon02}. Within reasonable spectral index limits this assumption has no significant effect on 
the derived luminosity functions.

We then evaluate  $S_{\nu}$ and $m_{i}$ using Equations
\ref{eq:absmag} and \ref{eq:radiopower} at
a series of redshifts ($\Delta z =0.005$) and find the
minimum and maximum redshifts where the source satisfies both the
optical and radio selection criteria as well as the limits of the redshift range being analysed (i.e. the appropriate redshift bin boundaries).
We can then calculate the integrated co-moving volume over which the source could have been
observed i.e. $V_{\rm max}$.

\subsection{The bivariate luminosity function}
In Figure \ref{fig:blf} we have plotted the bivariate luminosity function for all radio galaxies (including HERGs, LERGs and star-forming)  constructed
using Equation \ref{eq:lf} in three redshift bins. The left panel is the
luminosity function from our `local' redshift bin: $0.005 < z <
0.3$. It can be seen that the number density of radio galaxies rises steeply with
decreasing radio luminosity. There is a significant contribution to the
radio galaxy density from objects at faint optical magnitudes and low
radio luminosities. There is a large variance between the bivariate luminosity function bins for these
objects. Much of this contribution comes from local radio galaxies where the
radio emission arises from star-formation rather than from an
AGN (see right-hand column of Figure \ref{fig:blf_type}). The noise is due to the small volume (and hence absolute numbers)
in which these object satisfy the selection limits.
\begin{figure*}\setcounter{figure}{3}
  \begin{minipage}\textwidth
    \includegraphics[width=4.4cm, angle=90, trim=0 0 0 0]{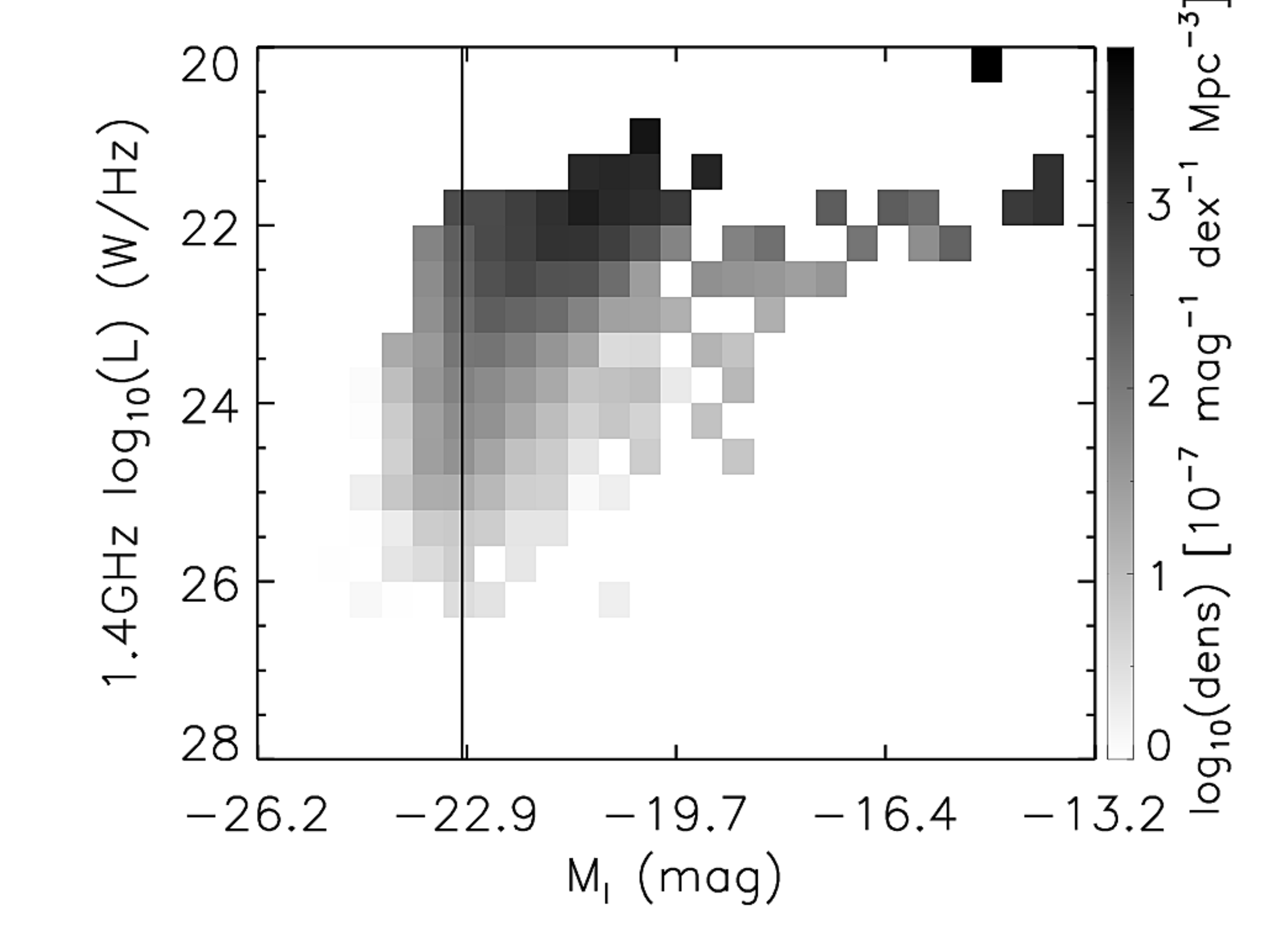}
    \includegraphics[width=4.4cm, angle=90, trim=0 0 0 0]{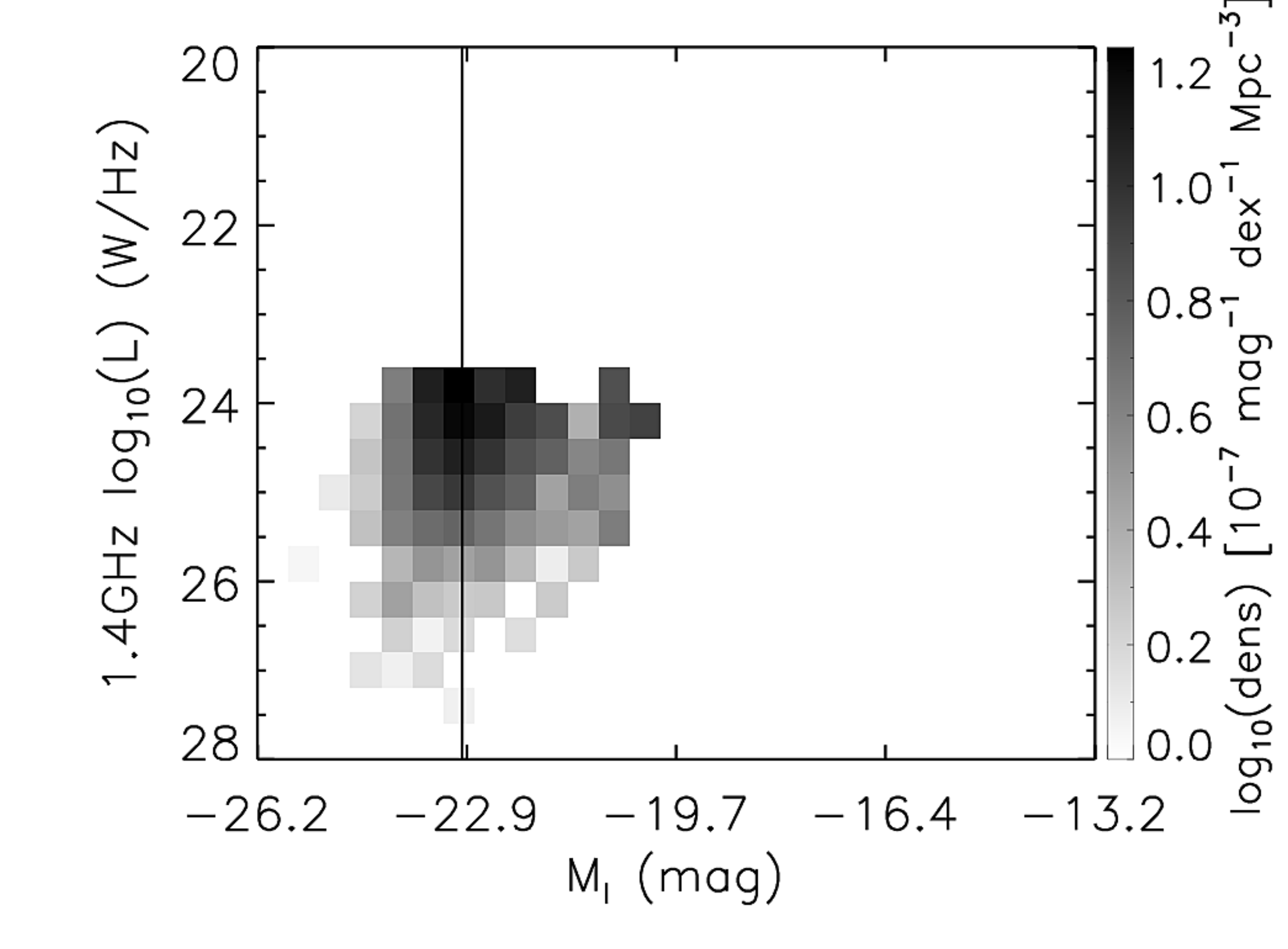}
    \includegraphics[width=4.4cm, angle=90, trim=0 0 0 0]{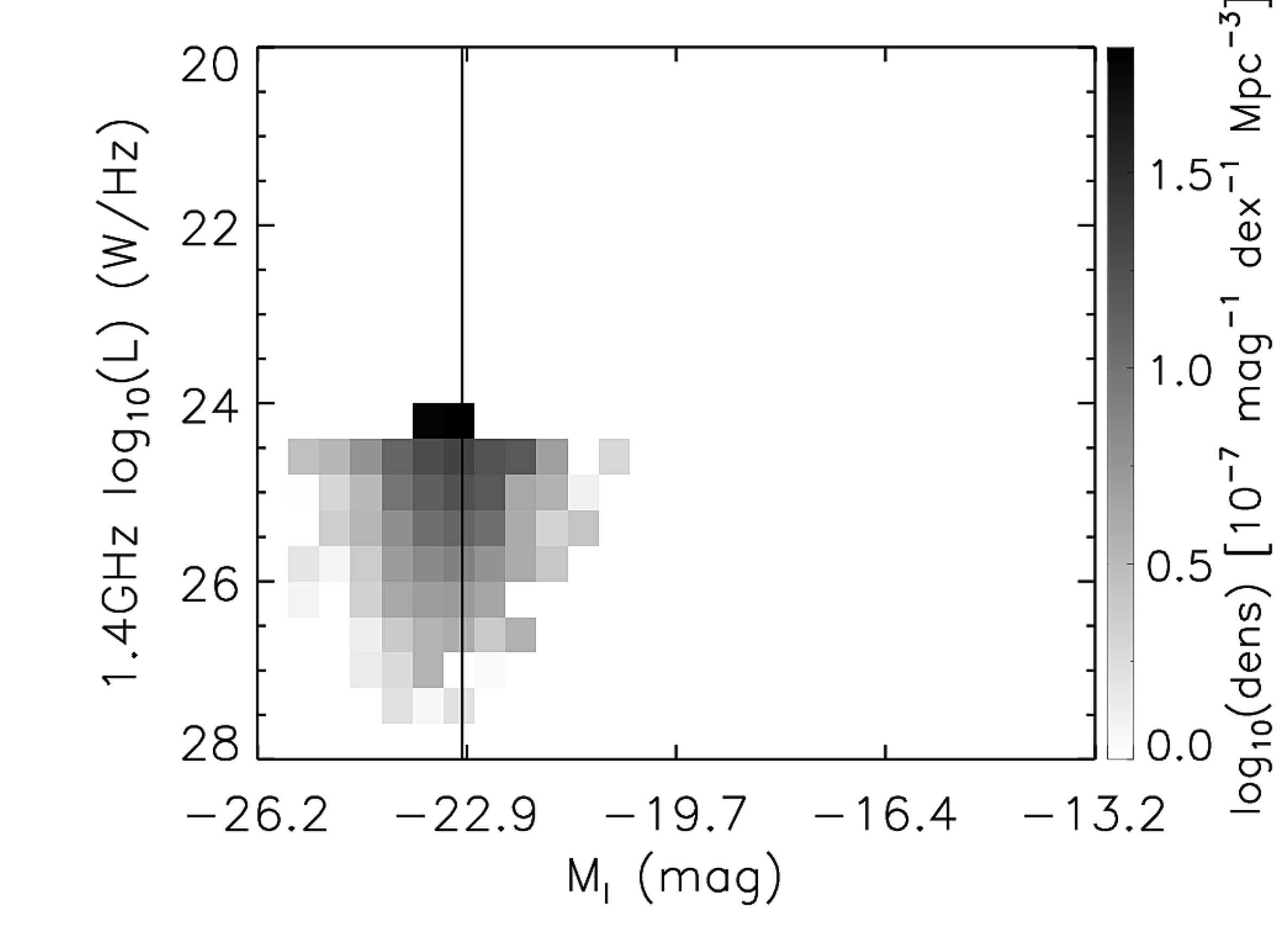}
\end{minipage}
\caption{\label{fig:blf} The bivariate luminosity
  distribution. The space density represented by a colour-scale in the
  optical magnitude--radio luminosity plane. The panels
  are for the different redshift bins: {\it left:}
  $0.005 < z < 0.30$; {\it middle:} $0.30 < z  <0.50$; and {\it
    right:} $0.5 < z < 0.75$. The flux-density and apparent magnitude selection limits
  correspond to progressively brighter radio and optical
  luminosity limits with increasing redshift. The absolute magnitude limit
  applied when analysing the redshift evolution of the radio
  luminosity function is shown as the vertical line. This corresponds
to the absolute magnitude limit at the high-redshift edge of the
highest redshift bin not inclusive of an  e-correction.}
\end{figure*}
The middle and right panels show the bivariate luminosity function for
redshift bins $0.3 < z < 0.5$ and $0.5 < z < 0.75$, respectively. As a
consequence of the flux density limits in both the optical and radio,
the luminosity function is restricted to progressively brighter (in
both optical luminosity and radio luminosity)
galaxies in the higher redshift bins. In all redshift bins the space density 
increases with decreasing radio luminosity. 

Later, when we consider redshift 
evolution in the radio luminosity function we will restrict our analysis to a subset of
galaxies with $M_{i}<-23$ (corresponding to the faintest galaxies we can detect at z=0.75). This 
limit is over-plotted as a vertical black line on Figure \ref{fig:blf}.

\subsection{The local radio luminosity function}
In order to produce a radio luminosity function we marginalize the bivariate
luminosity function over optical luminosity. That is, we calculate the
space density of radio galaxies, per radio luminosity bin, integrated over
all observed optical luminosities. In Figure \ref{fig:local} we produce the
local radio galaxy luminosity function. The radio luminosity
function for all radio galaxies is plotted as {\it open diamonds} and
the luminosity functions separated for AGN and star forming galaxies are plotted as
{\it open triangles} and {\it open squares}, respectively. Below a
radio luminosity of $\sim L_{\rm 1.4\,GHz} \sim 10^{23} {\rm W~Hz}^{-1}$ the
star-forming galaxies dominate the space
density of sources, while above this luminosity the AGN population
dominates. These luminosity functions are also tabulated in Table \ref{tab:local_all} 
\begin{figure}\setcounter{figure}{4}
      \includegraphics[width=6.4cm, angle=90]{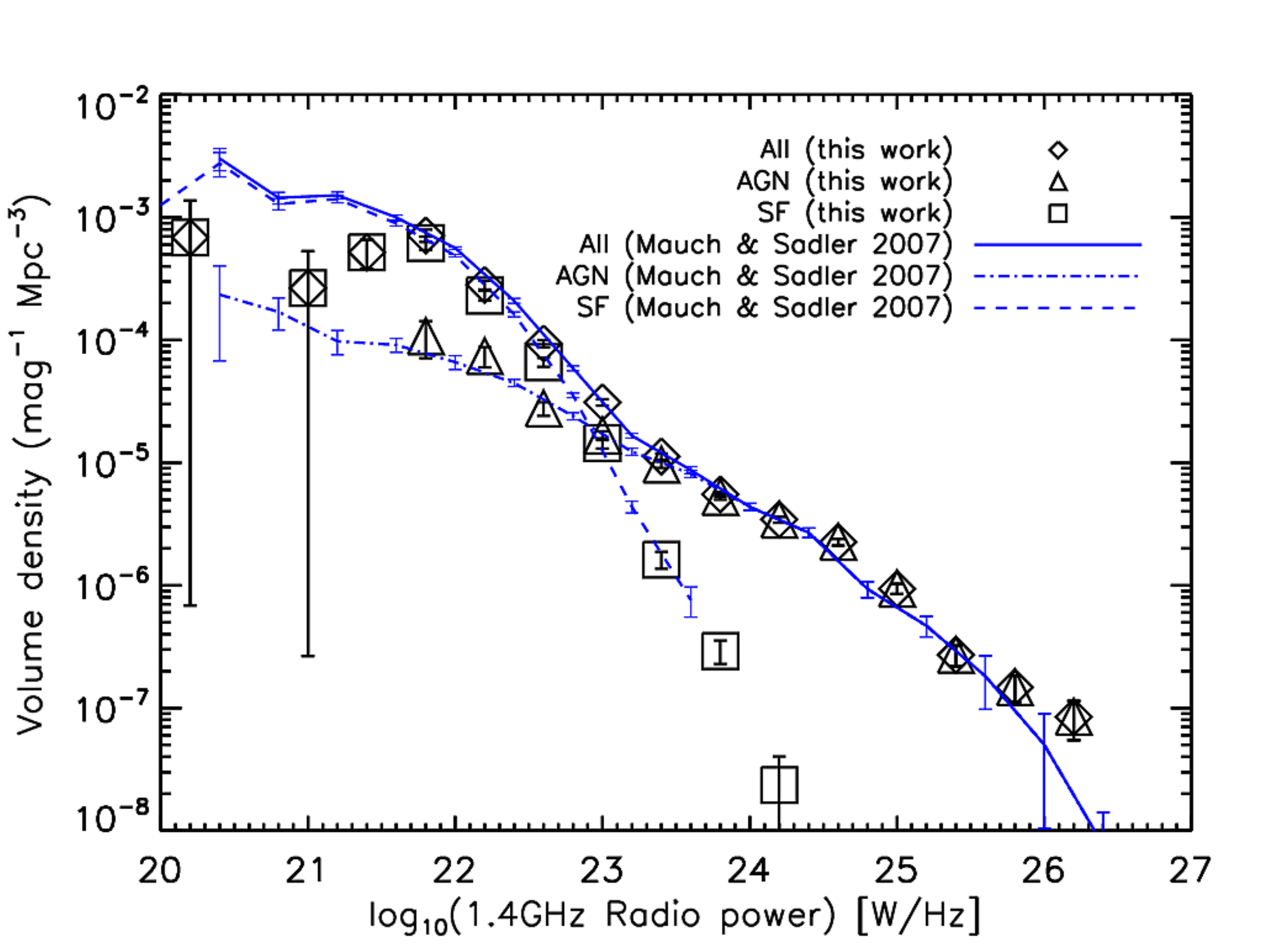}
\caption{\label{fig:local} The radio  luminosity function for 
  our $0.005 < z < 0.3$ redshift bin ({\it diamonds}). Also, plotted
  the luminosity function separated into radio AGN ({\it triangles}) and star-forming radio
  galaxies ({\it squares}). The star-forming galaxies outnumber the
  AGN below $P_{\rm 1.4\,GHz} \sim 10^{23}{\rm W~Hz}^{-1}$ but
  contribute very little to the number density at higher radio luminosities.
Also shown is the radio luminosity functions from \citet{mauch07} for
all sources
({\it solid blue line}), radio AGN ({\it dot-dashed blue line}) and
radio star-forming galaxies ({\it dashed blue line}).}
\end{figure}

Also in Figure \ref{fig:local} we overplot {\it as blue
  lines} the local radio luminosity function of 6dFGRS galaxies from
\citet{mauch07} for all radio galaxies ({\it solid line}), AGN ({\it
  dot-dashed line}) and star-forming galaxies ({\it dashed line}). Our
luminosity functions and those of \citet{mauch07} are in good  agreement and the local radio
luminosity function of \citet{mauch07} is known to be in good agreement with determinations
measured by other authors \citep{machalski00,sadler02,mao12,best12}. It should be noted that the
comparison of these luminosity functions is not expected to be exact.
It is clear from the bivariate luminosity functions in Figure
\ref{fig:blf} that the space density of radio galaxies measured will depend
on the range of optical luminosities sampled. The fainter our optical limits the
more galaxies at a given radio luminosity we will find. Strictly then,  to compare radio
luminosity functions we should integrate over the same optical
magnitude range to ensure consistency.  For local luminosity functions this effect should be
small, since at the lowest redshifts all surveys will probe down to very faint optical magnitudes 
(albeit noisily because of the reducing volume) and the $1 /\ V_{\rm max}$ correction will return approximately the 
same luminosity function. However, when comparing luminosity functions at different redshifts it
 is important to use the same range of optical luminosities, as demonstrated in Figures \ref{fig:blf} and  \ref{fig:blf_type}.
\begin{table*}
\caption{The local ($0.005<z<0.3$) 1.4\,GHz radio luminosity function  from this sample separated into
  radio AGN and star-forming galaxies. The radio AGN are further separated into LERGs and HERGs.}
\begin{center}
\begin{tabular}{ccccccccccccc}
\hline\hline
                             &                  & \hspace{-1.3cm} All  galaxies &                 &    \hspace{-1.3cm}  SF  galaxies  &                & \hspace{-1.3cm}  Radio AGN           &          & \hspace{-1.3cm}  LERGs   &        & \hspace{-1.3cm}  HERGS               \\
log$_{10} P_{1.4} $   &    N            &  $\log(\Phi)$                           &      N        &       $\log(\phi)$                           &        N          &        $\log(\Phi)$              &    N    &             $\log(\Phi)$           &    N    &        $\log(\Phi)$     \\
  (W Hz$^{-1}$)       &                  &   (mag$^{-1}$ Mpc$^{-3}$)       &                 &         (mag$^{-1}$ Mpc$^{-3}$)        &                      &   (mag$^{-1}$ Mpc$^{-3}$) &           &         (mag$^{-1}$ Mpc$^{-3}$)          &        &   (mag$^{-1}$ Mpc$^{-3}$)    \\ \hline
 21.00                   &     1             &   $-3.58^{+0.30}_{-3.00}$         &       1         &    $-3.58^{+0.30}_{-3.00}$                &                     &                                            &          &                         &                 &                                           \\
 21.40                  &      14         &      $-3.28^{+0.10}_{-0.13}$       &        14     &       $-3.28^{+0.10}_{-0.13}$               &                     &                                             &        &                          &                 &                                      \\     
 21.80                    &    76            &   $-3.15^{+0.05}_{-0.05}$        &       67     &          $-3.21^{+0.05}_{-0.06}$            &        9           &     $3.97-^{+0.12}_{-0.18}$      &     8       &  $-4.00^{+0.13}_{-0.19}$          &    1          &      $-4.75^{+0.30}_{-3.00}$           \\         
 22.20                   &     156        &      $-3.55^{+0.03}_{-0.04}$       &     128       &            $-3.64^{+0.04}_{-0.04}$        &       28          &     $-4.13^{+0.08}_{-0.09}$    &     22       &  $-4.19^{+0.08}_{-0.10}$           &    6        &       $-5.06^{+0.15}_{-0.23}$            \\       
 22.60                   &     208         &   $-4.03^{+0.03}_{-0.03}$        &     144        &            $-4.18^{+0.03}_{-0.04}$        &       64          &     $-4.56^{+0.05}_{-0.06}$   &      46       &  $-4.72^{+0.06}_{-0.07}$           &    18       &        $-5.07^{+0.09}_{-0.12}$          \\  
 23.0                   &     255        &    $-4.51^{+0.03}_{-0.03}$        &       109      &            $-4.84^{+0.04}_{-0.04}$          &       146        &     $-4.78^{+0.03}_{-0.04}$   &       124    &  $-4.85^{+0.04}_{-0.04}$           &     22          &     $-5.58^{+0.08}_{-0.10}$          \\   
  23.40                   &    314          &    $-4.95^{+0.02}_{-0.03}$        &     41        &          $-5.79^{+0.06}_{-0.07}$          &       273        &     $-5.02^{+0.03}_{-0.03}$   &       234    &   $-5.09^{+0.03}_{-0.03}$          &    39           &     $-5.84^{+0.06}_{-0.08}$          \\ 
 23.80                   &     440        &      $ -5.26^{+0.02}_{-0.02}$      &      21          &        $-6.54^{+0.09}_{-0.11}$          &       419        &     $-5.28^{+0.02}_{-0.02}$    &      368    &  $-5.34^{+0.02}_{-0.02}$            &     51          &    $-6.15^{+0.06}_{-0.07}$           \\  
  24.20                   &    349          &     $-5.46^{+0.02}_{-0.02}$       &       2       &          $-7.63^{+0.23}_{-0.53}$           &       347        &     $-5.47^{+0.02}_{-0.02}$      &     319   &  $-5.52^{+0.02}_{-0.02}$             &    28          &    $-6.43^{+0.08}_{-0.09}$          \\ 
 24.60                   &     242         &      $-5.65^{+0.03}_{-0.03}$      &                 &                                                      &        242        &     $-5.65^{+0.03}_{-0.03}$       &    217   &   $-5.71^{+0.03}_{-0.03}$           &     25         &      $-6.52^{+0.08}_{-0.10}$           \\
25.00                    &     112         &        $-6.03^{+0.04}_{-0.04}$     &                 &                                                     &        112        &     $-6.03^{+0.04}_{-0.04}$       &    101   &   $-6.07^{+0.04}_{-0.05}$          &     11      &      $-7.04^{+0.11}_{-0.16}$              \\
 25.40                    &     27           &      $-6.57 ^{+0.08}_{-0.09}$        &                 &                                                   &         27         &     $-6.57^{+0.08}_{-0.09}$        &    20    &   $-6.78^{+0.09}_{-0.11}$          &      7        &     $-6.98^{+0.14}_{-0.21}$            \\ 
 25.80                    &     17            &     $-6.83^{+0.09}_{-0.12}$      &                 &                                                     &         17         &     $-6.83^{+0.09}_{-0.12}$        &     7     &   $-7.26^{+0.14}_{-0.21}$         &       10        &     $-7.03^{+0.12}_{-0.16}$          \\ 
 26.20                   &       8          &     $-7.07^{+0.13}_{-0.19}$       &                 &                                                       &          8          &     $-7.07^{+0.13}_{-0.19}$       &      3     &   $-7.42^{+0.20}_{-0.37}$         &        5         &    $-7.33^{+0.16}_{-0.26}$           \\   

\hline
 
\end{tabular}
\end{center}
\label{tab:local_all}
\end{table*}

\subsection{Separating the luminosity function of radio AGN by accretion mode}
As described in Section \ref{sec:class} the AGN in our sample have been
classified as either LERGs or HERGs based on their optical spectral properties. 
In Figure \ref{fig:blf_type} we show the bivariate luminosity function for LERGs (left column), HERGs (middle column) and 
star forming radio galaxies (right column) in three redshift bins (top to bottom). The star forming radio galaxies are shown only for the low
redshift bin, since there are no such objects in the catalogue with higher redshifts.  

In Figure \ref{fig:rlftype} (and tabulated in Table \ref{tab:local_all})
we show the radio luminosity functions for LERGs and HERGs separately, for
AGN in our  `local' $0.005 < z < 0.3$ redshift bin.  We have summed over
 optical luminosities down to the limit of our sample ($m_{i} < 20.5$). The local LERG and HERG
luminosity functions of \citet[][parameters taken from \citealt{heckman14}) are also shown ({\it solid lines}]{best12}. The LERG
luminosity functions are similar in shape and normalisation. The
HERG luminosity functions, however, are significantly different. The
HERG luminosity function presented here has a higher  space density,
especially at low radio luminosities. As noted earlier, we don't expect
radio luminosity functions to agree unless they sample the same range
of optical luminosities, however, when measured locally this effect should be small and cannot explain the discrepancy. 

In the case of the HERGs there are also differences in the classification techniques. \citet{best12} were deliberately strict in 
removing star-forming radio  galaxies. This conservative approach is well justified since at the faint-end of the radio luminosity function the 
star-forming radio galaxies can dominate in number density over the AGN by approximately an order of magnitude (see Figure \ref{fig:local}). In this case 
mis-classifying a small fraction of star-forming galaxies as AGN can have a considerable effect on the measured space density of AGN.

Based on comparison of common objects, some radio galaxies classified as star-forming radio galaxies by \citet{best12} are expected
to be classified as HERGs in this sample \citep{ching15}. \citet{best12} use a combination of three tests to  remove radio galaxies
 where the radio emission is suspected to arise from star-formation: a standard BPT emission line diagnostic; a ratio of radio-to-emission-line luminosity; and a method using the
strength of the 4000\AA\, break (measured via the D4000 index)  and the ratio of radio luminosity to stellar mass. While \citet{ching15} used methods similar to the first two --- they do not apply the third. The reason for
this is two-fold. Firstly, some of the spectra (such as those derived from the WiggleZ survey) have low continuum signal-to-noise and poor correction of the wavelength response -- meaning the 4000\AA\, 
break strength cannot be accurately measured. Secondly, while \citet{best12} include objects classified as QSOs, they only do so if they were targeted as galaxies (i.e. not point-sources). In contrast the catalogue of \citet{ching15} includes  broad-line and point-like objects. A test involving the D4000 strength is inappropriate for such objects since they will have strong blue continuum.

In addition, while a D4000 test is well justified for the LERGs which are a well separated population having strong D4000 strengths well above the star forming cut  --- this is not true for the narrow line HERGs.  Radio galaxies classified as HERGs using a BPT diagnostic have a significant fraction of the population within the star-forming region.  These are likely HERGs incorrectly classified as star forming galaxies using this criterion \citep[see {\it left middle panel} of Figure 9 in][]{best05}. As another example, \citet{herbert10}  measure the D4000 strength for a population of high power HERGs where the radio emission clearly arises from AGN activity. This population has an approximately uniform distribution of D4000 strengths between $\sim$ 1.1--1.7. While most of these galaxies would be classified as AGN because of their high radio power, since the stellar population will fade on time scales much longer than the time scale on which radio activity can change such objects can move into an area of parameter space where they would be mis-classified as HERGs. 
While many of the radio AGN misclassified as star-forming using this test can be correctly classified based on the other two test \citep[see][Appendix A2]{best12} these examples again demonstrate the difficulty in separating HERGs and star-forming radio galaxies.

In Figure \ref{fig:herg_sf_compare} we demonstrate that the differences between  our local HERG luminosity function and that of \citet{best12} can be explained by differences in the classifications in the star-forming objects (at low radio powers) and the inclusion of broad-line objects (at $L_{1.4\rm{GHz}}\sim 10^{24}$\,W\,Hz$^{-1}$). In the top panel of Figure \ref{fig:herg_sf_compare} we show the local luminosity function of the star-forming radio galaxies (dotted lines) and HERGs (dashed lines) for our sample (blue) and that of \citet{best12} (red). The solid lines show the sum of the HERGs and star-forming luminosity functions.  The luminosity functions of the sum are well matched at the faint-end between the two samples. The difference being some objects classified as HERGs by \citet{ching15} were removed as star-forming galaxies by \citet{best12}. 

There is, however, a difference at $L_{1.4\rm{GHz}}\sim 10^{24}$\,W\,Hz$^{-1}$ with the number density of objects in our sample being higher by a factor of $\sim 2$. In the bottom panel we show the same luminosity functions but with the broad-line objects removed from our HERG luminosity function. The two determinations are in much better agreement and the remaining difference is not statistically significant. 

The determination of the true HERG luminosity function at low radio power is a fundamentally difficult problem stemming from the fact that the 
number density of star-forming radio galaxies is so high in this regime. A faultless classification would likely require high spatial resolution radio continuum imaging. 
However, it  is not critical in the context of this paper i.e. measuring the redshift evolution of the luminosity function. This is 
because the radio luminosity limits in our higher redshift bins (set by our flux limit) are outside the regime where star-formation powered radio galaxies are important. For 
example, even in our intermediate redshift bin we are restricted to radio luminosities  $L_{1.4\rm{GHz}}\gtrsim 10^{24}$\,W\,Hz$^{-1}$. Above these radio luminosities the number density of HERGs 
quickly rises above the star-forming galaxies, regardless of the classification method used.

\begin{figure*}\setcounter{figure}{5}
  \begin{minipage}\textwidth
    \includegraphics[width=4.2cm, angle=90, trim=0 0 0 0]{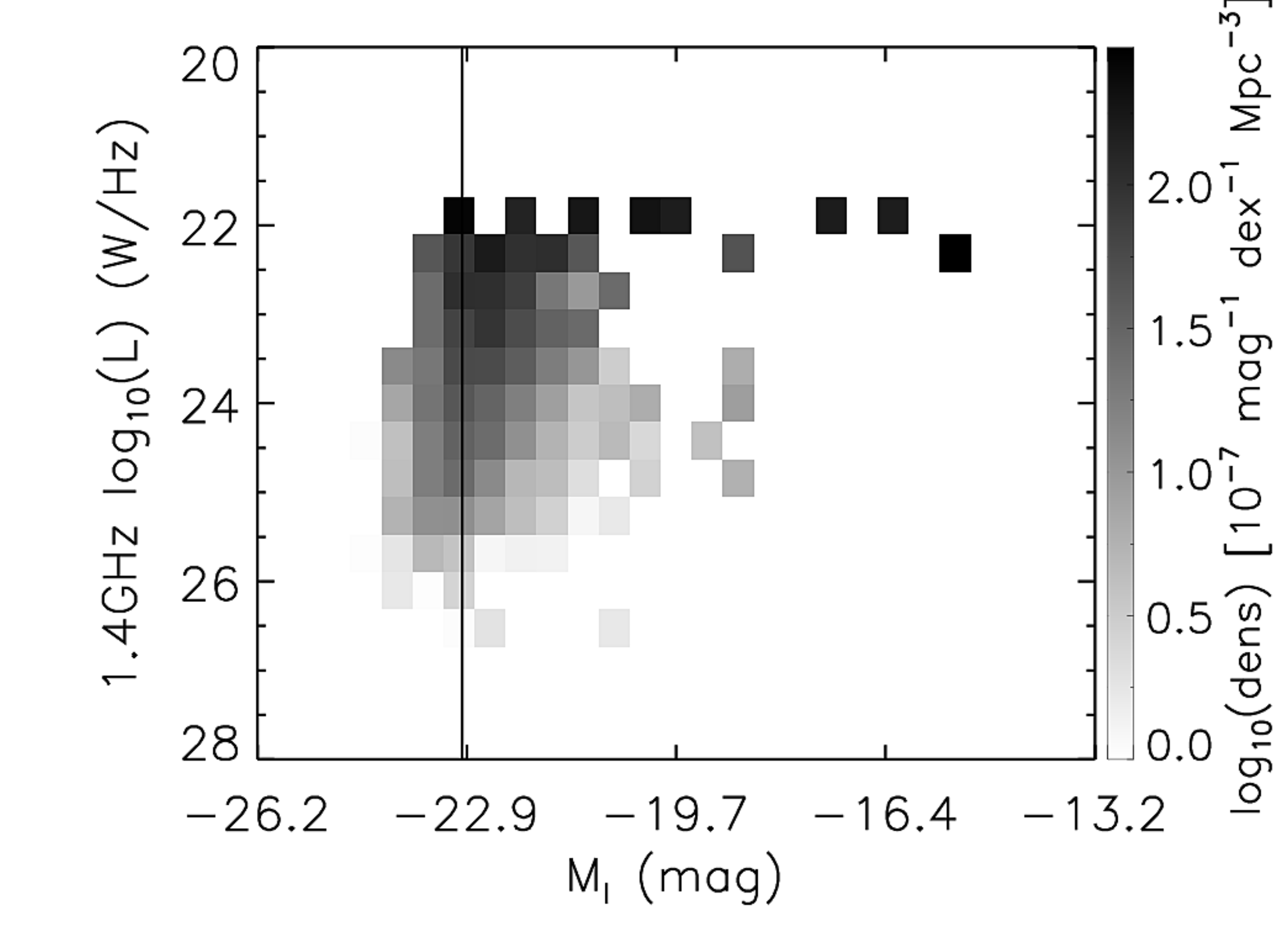}
    \includegraphics[width=4.2cm, angle=90, trim=0 0 0 0]{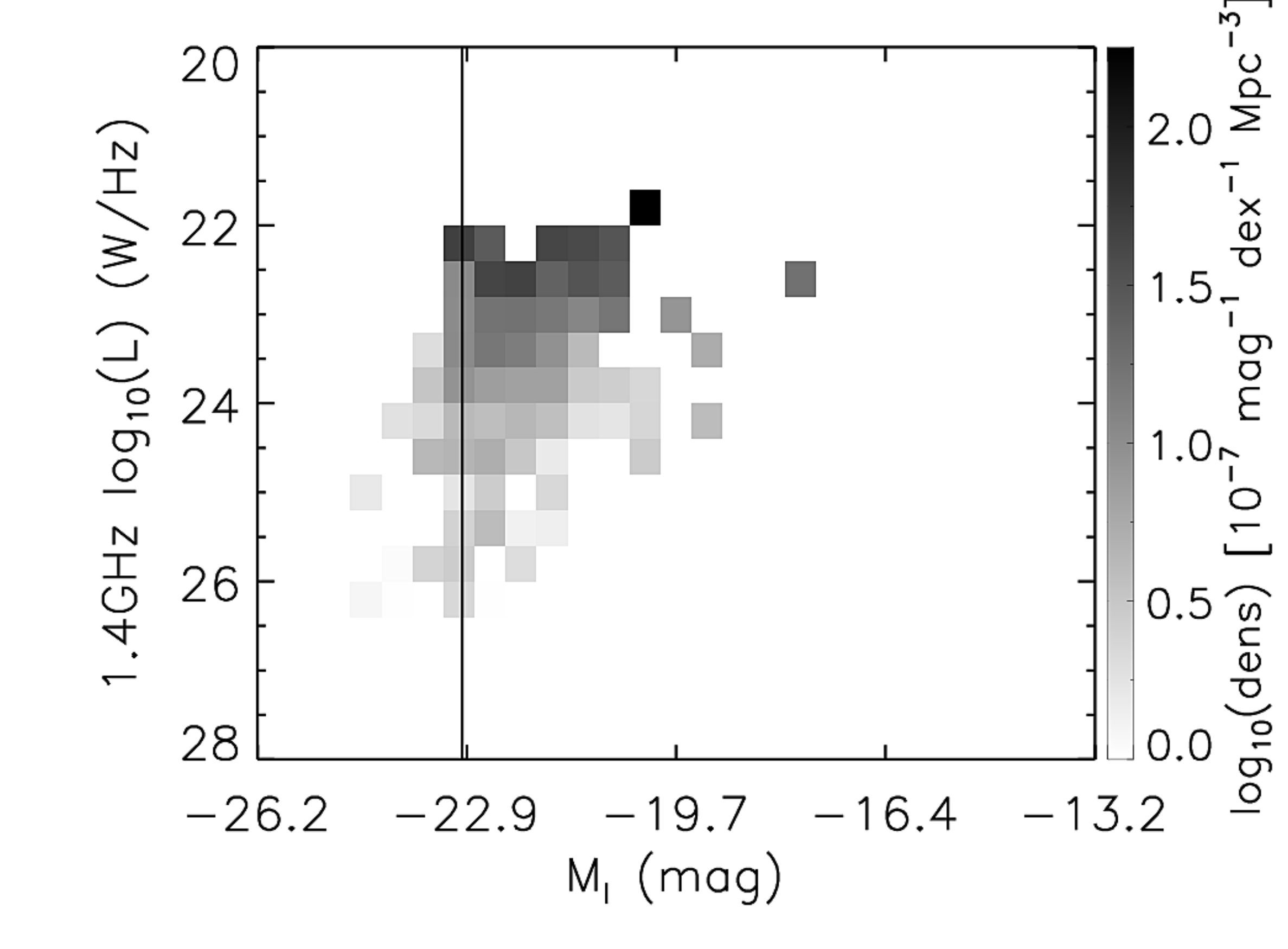}
    \includegraphics[width=4.2cm, angle=90, trim=0 0 0 0]{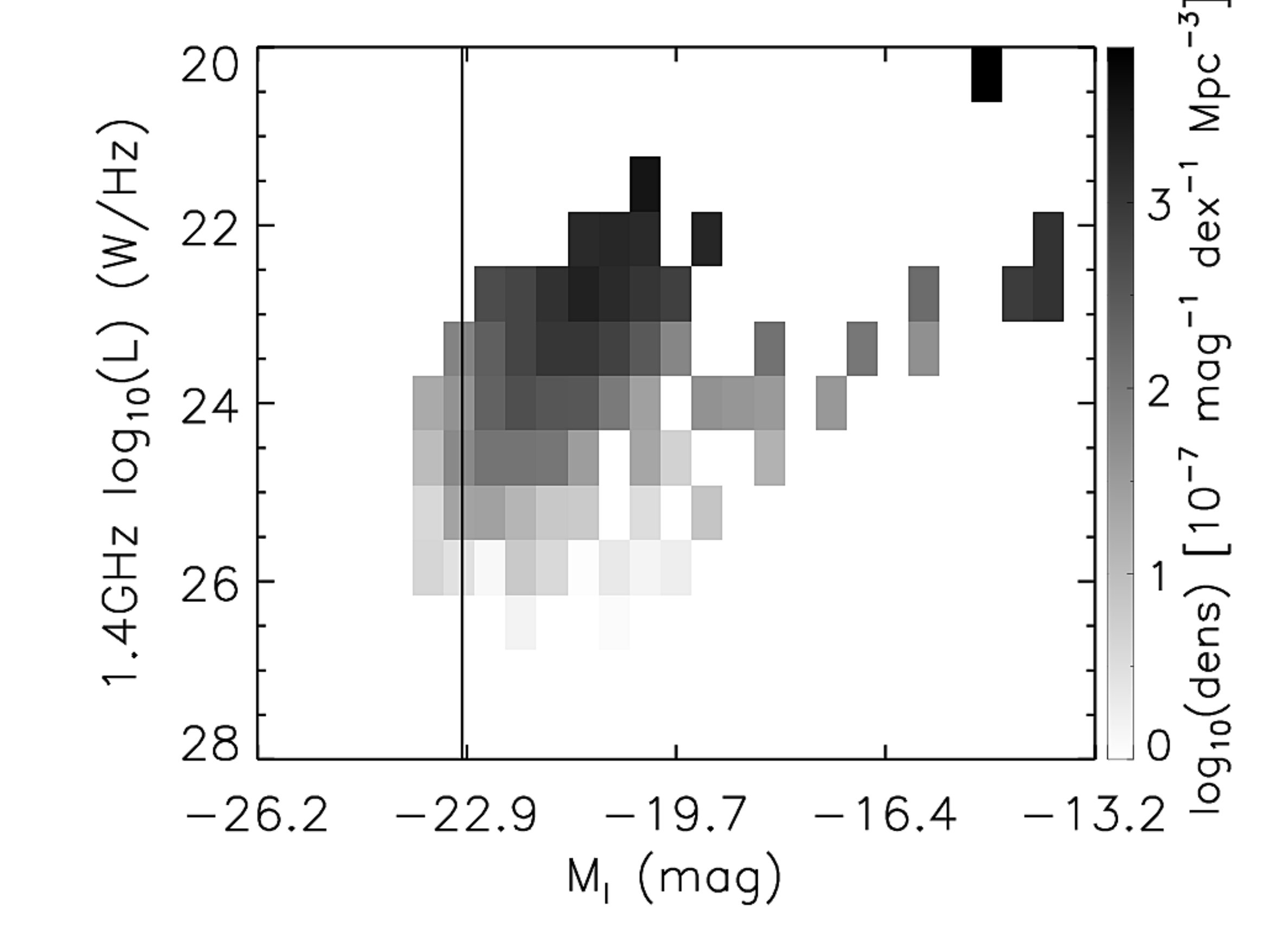}
  \end{minipage}
  \begin{minipage}\textwidth
    \includegraphics[width=4.2cm, angle=90, trim=0 0 0 0]{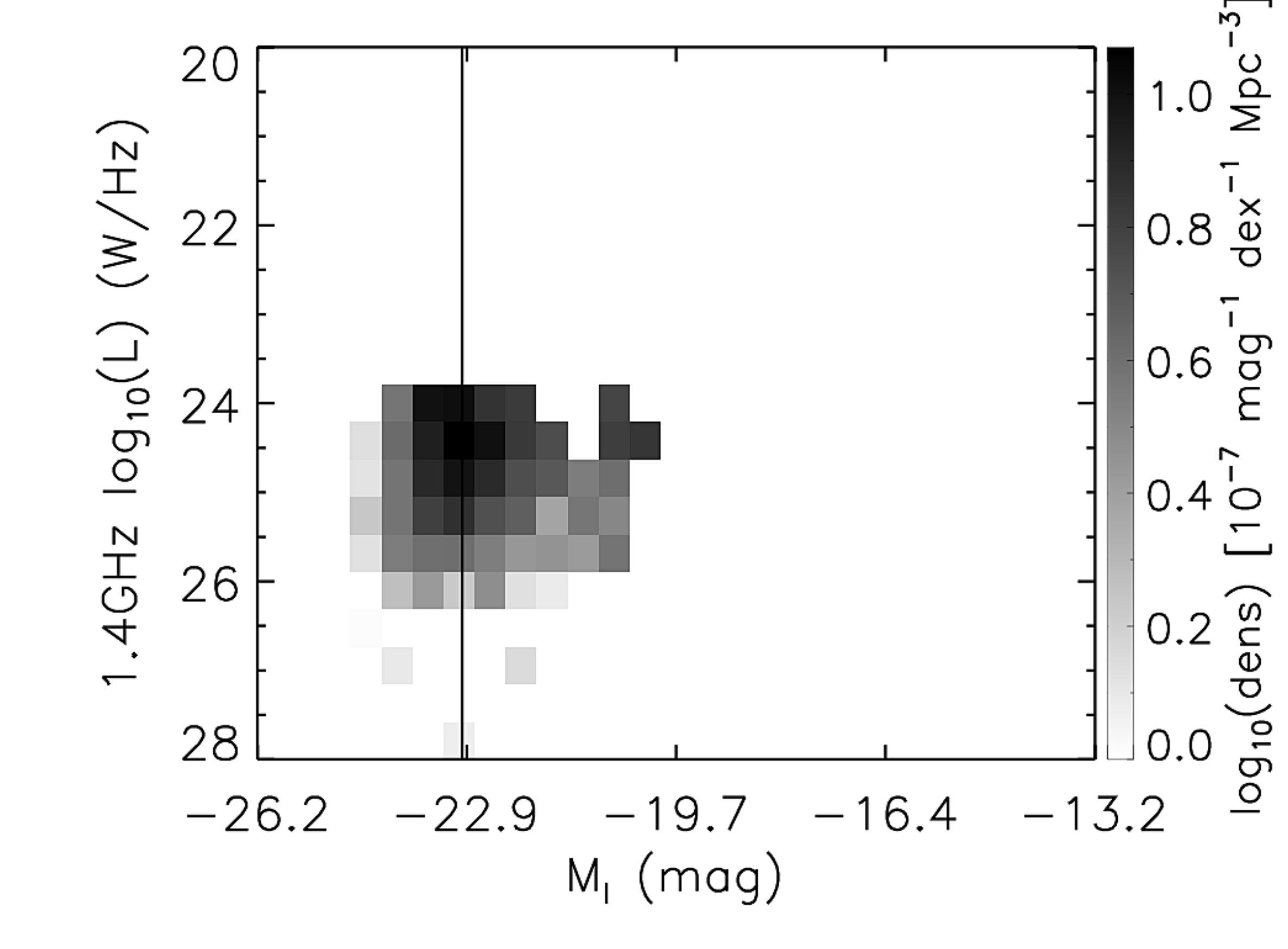}
    \includegraphics[width=4.2cm, angle=90, trim=0 0 0 0]{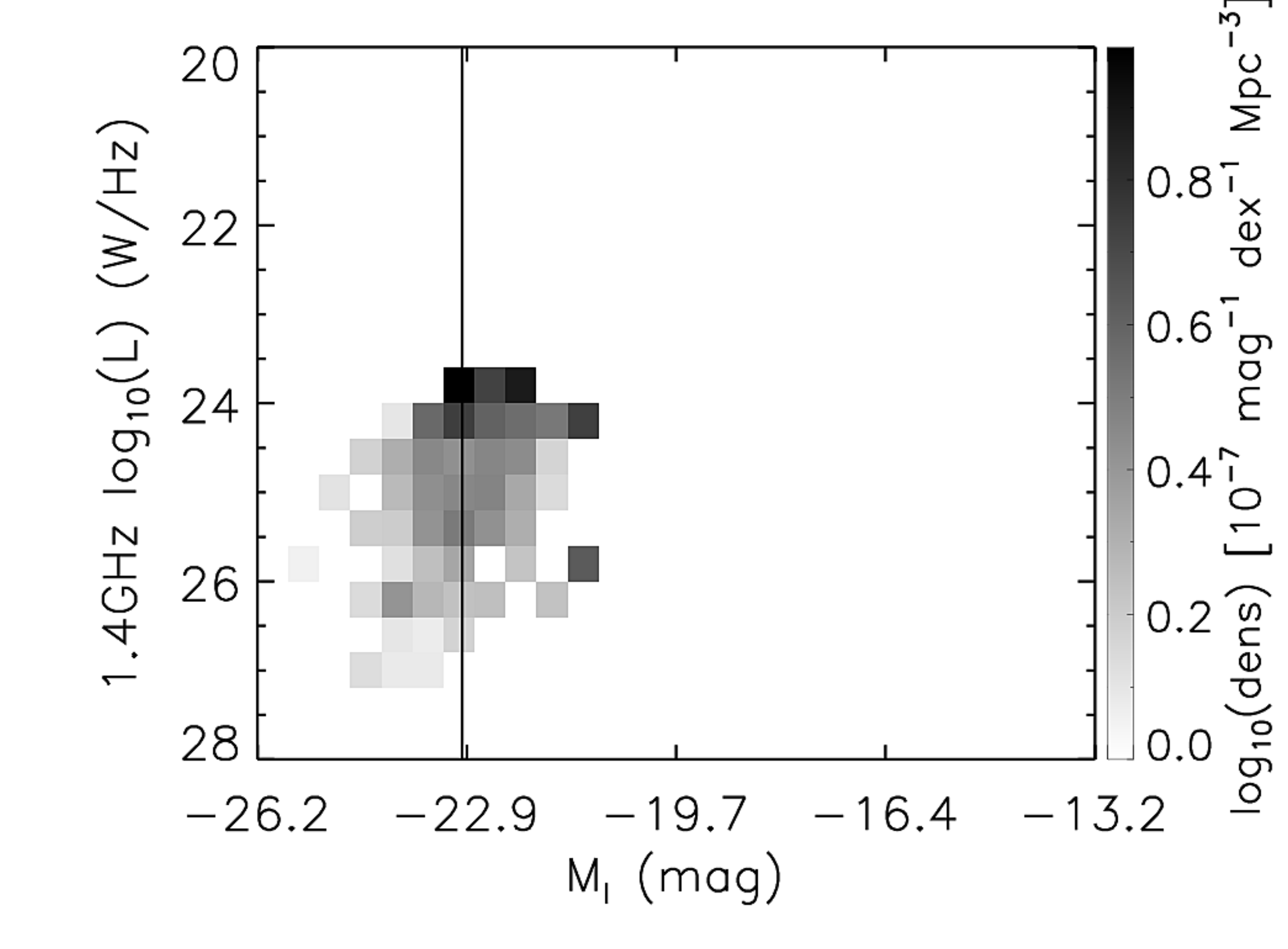}
  \end{minipage}
  \begin{minipage}\textwidth
    \includegraphics[width=4.2cm, angle=90, trim=0 0 0 0]{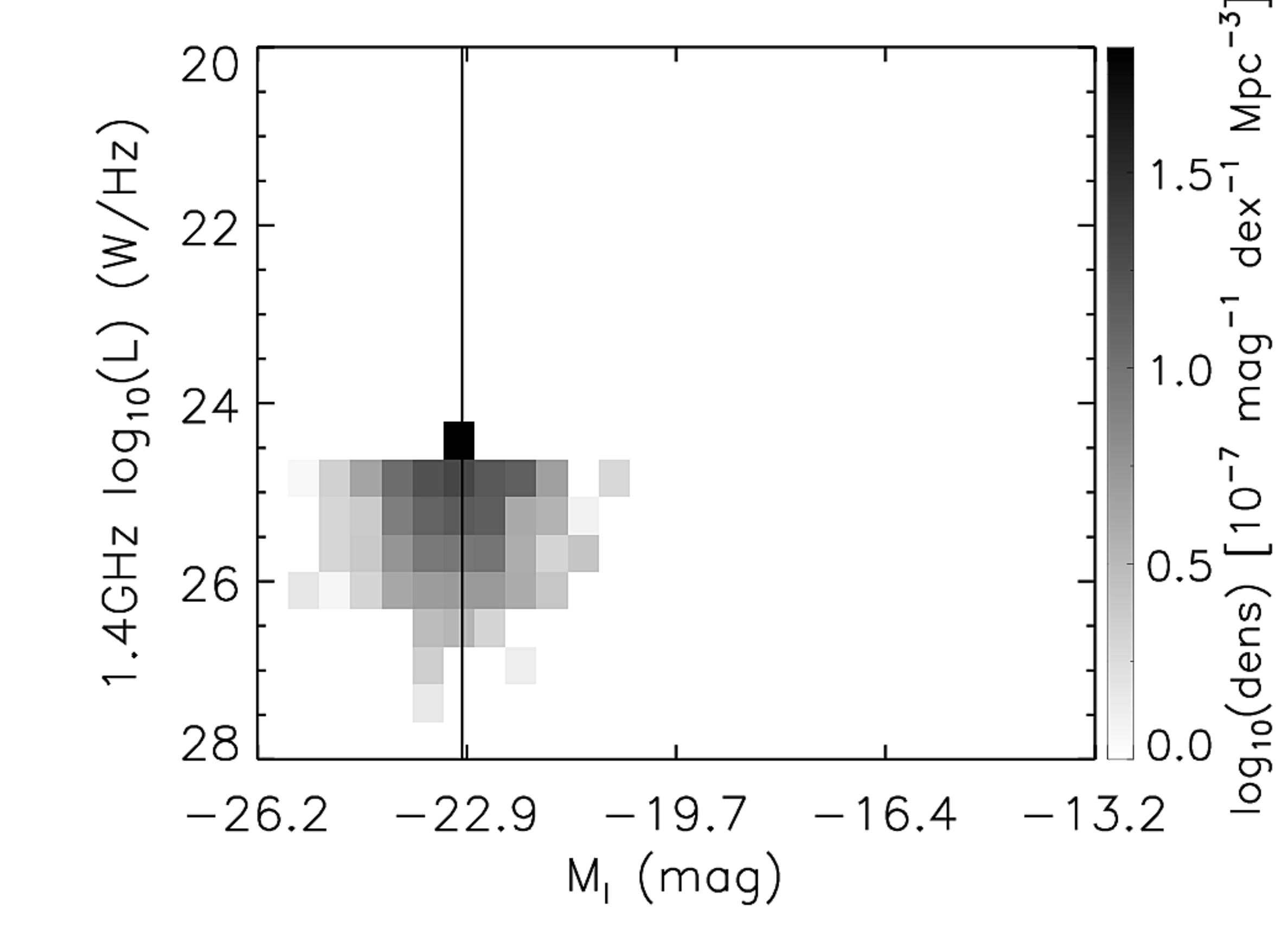}
    \includegraphics[width=4.2cm, angle=90, trim=0 0 0 0]{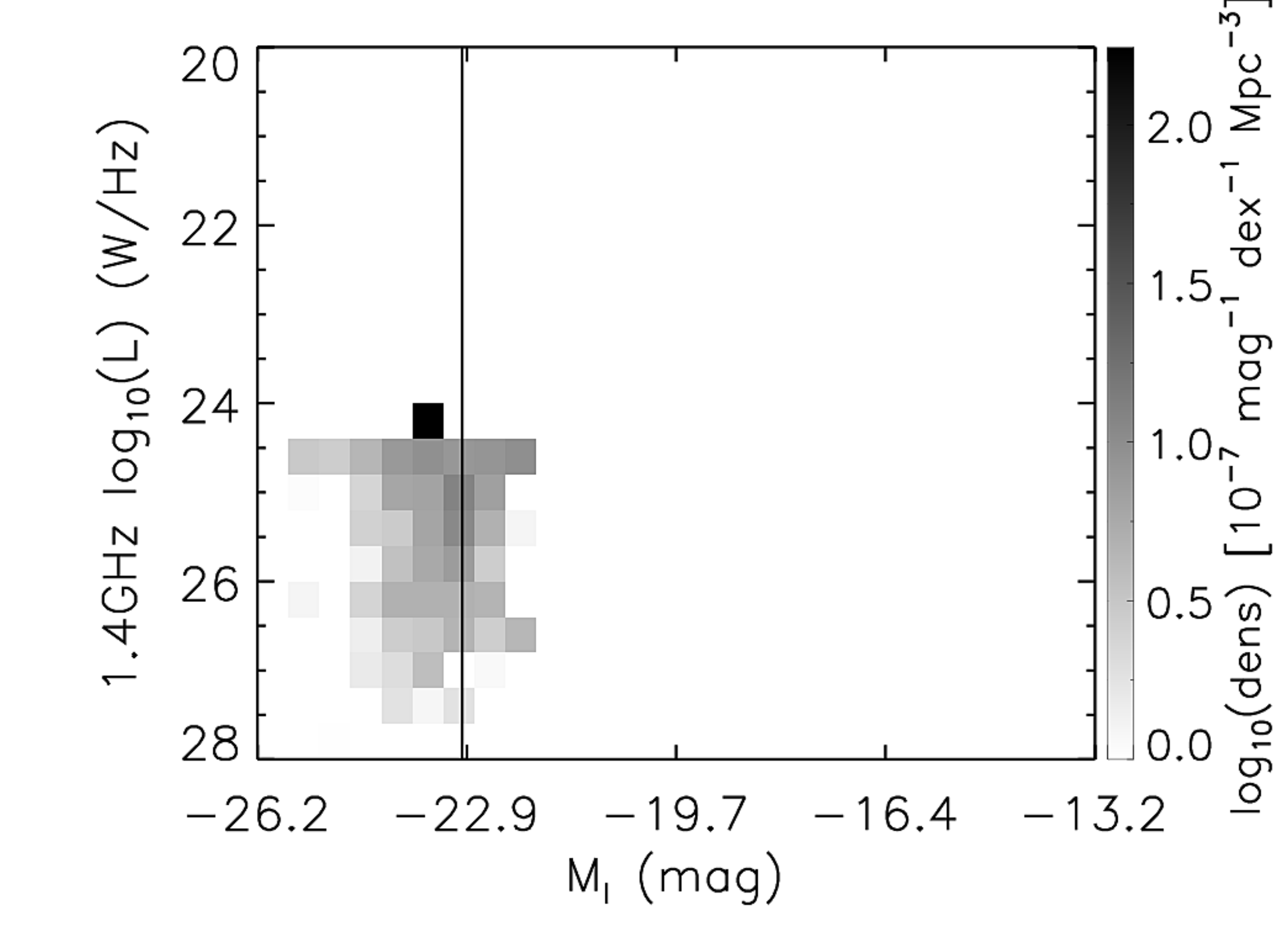}
  \end{minipage}
\caption{\label{fig:blf_type}The bivariate luminosity
  distribution for LERGs (left column), HERGS (middle column) and star forming radio galaxies (right column).  
The rows  are for the different redshift bins: {\it top row:}
  $0.005 < z < 0.30$; {\it middle row:} $0.30 < z  <0.50$; and {\it
    bottom row:} $0.5 < z < 0.75$. Only the lowest redshift bin is shown for the star-forming radio galaxies since there are no objects 
 with this classification in higher redshift bins.  The flux-density and apparent magnitude selection limits
  correspond to progressively brighter radio and optical
  luminosity limits with increasing redshift. The absolute magnitude limit
  applied when analysing the redshift evolution of the radio
  luminosity function is shown as the vertical line. This corresponds
to the absolute magnitude limit at the high-redshift edge of the
highest redshift bin.}
\end{figure*}

\begin{figure}\setcounter{figure}{6} 
   \includegraphics[width=6.2cm, angle=90]{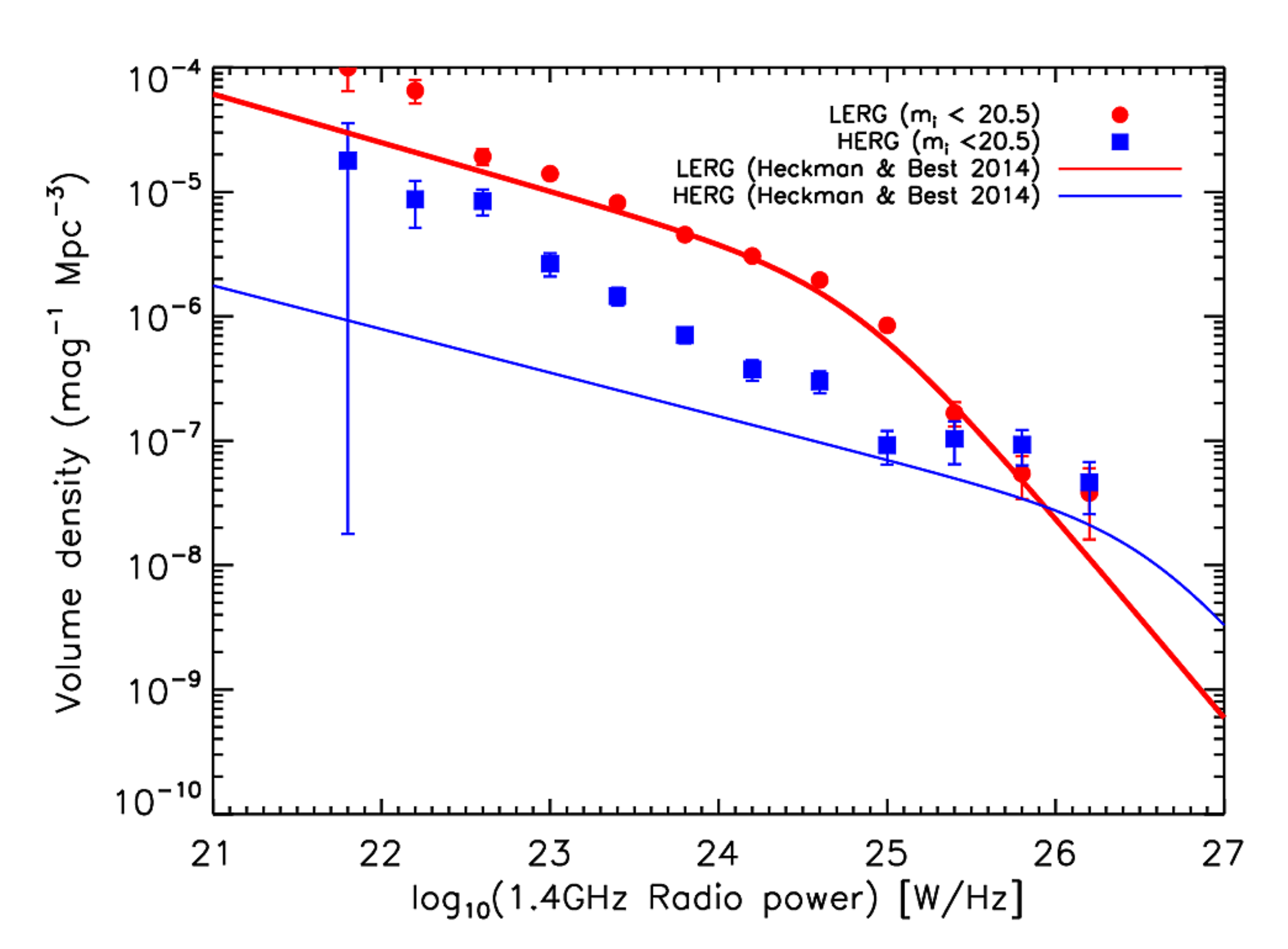}
\caption{\label{fig:rlftype} The radio luminosity function of AGN in
  our $0.005 < z < 0.3$ redshift bin separated into LERGs ({\it red  filled circles})
  and HERGs ({\it blue  filled squares })  summed over optical magnitudes down to our limit of  $i<20.5$. 
Over-plotted are the luminosity functions of local LERGS ({\it red solid line}) and HERGs ({\it blue solid line}) from \citet{heckman14}. }
\end{figure}

\begin{figure}\setcounter{figure}{7}
    \includegraphics[width=6.2cm, angle=90, trim=0 0 0 0]{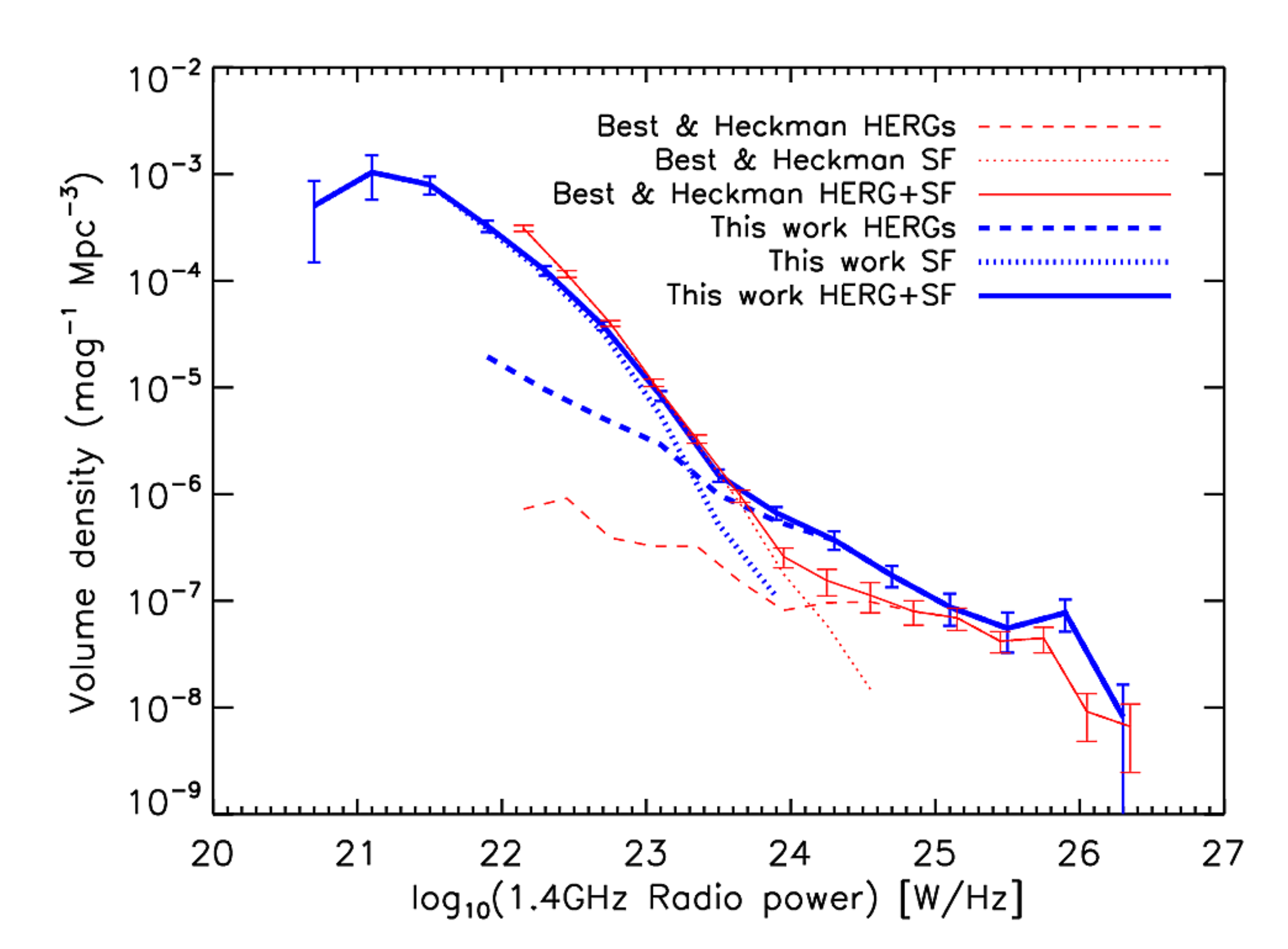}
    \includegraphics[width=6.2cm, angle=90, trim=0 0 0 0]{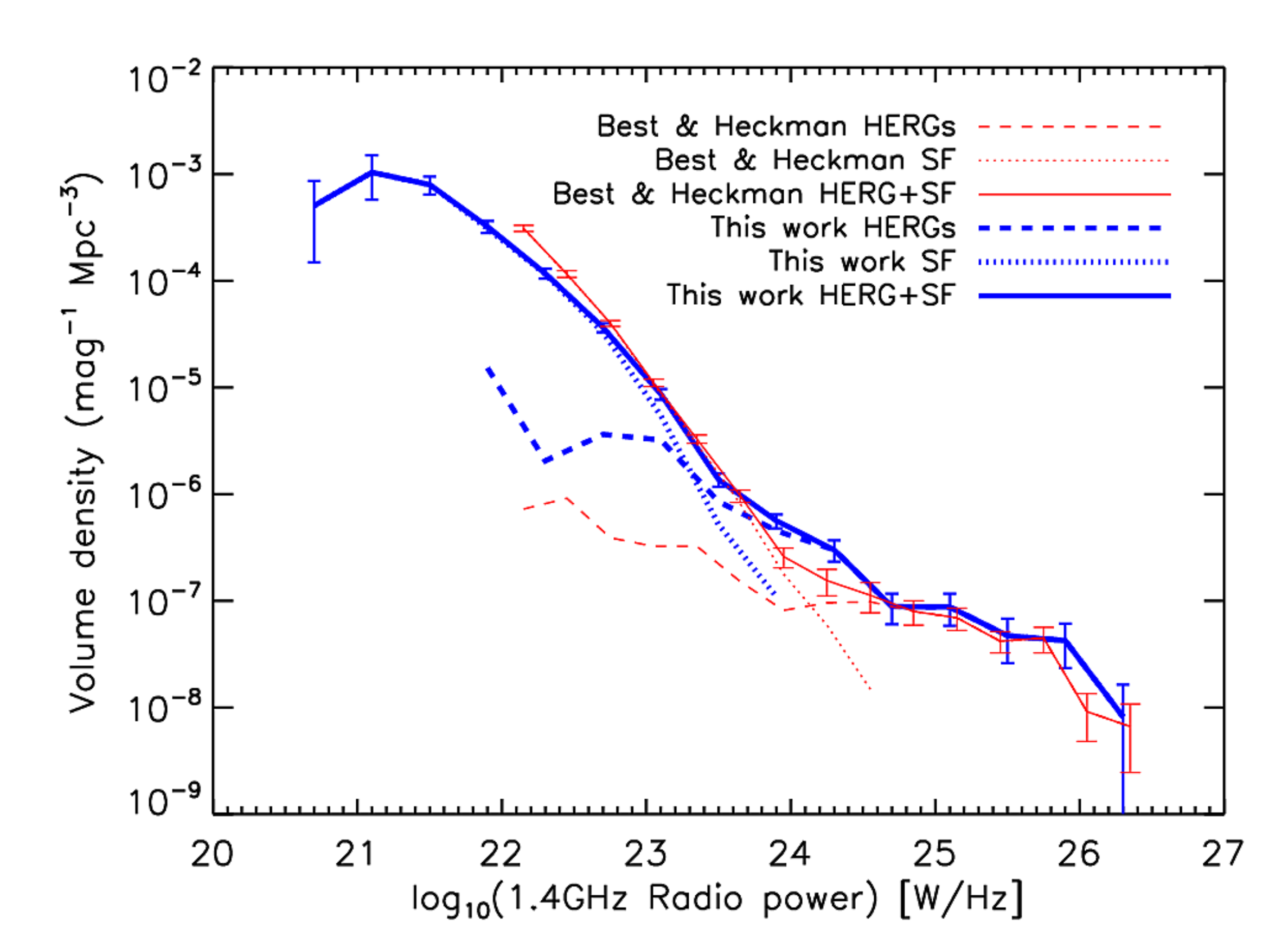}
\caption{\label{fig:herg_sf_compare}Top panel: the radio luminosity function of star forming radio 
galaxies ({\it blue dotted line}), HERGs ({\it blue dashed line}) and their sum ({\it blue solid line}) in our sample and those
from \citet{best12} ({\it red dotted, red dashed and red solid lines, respectively}). The bottom panel is the same as the top panel 
but with the broad line AGN removed from our HERG sample.}
\end{figure}

\subsection{Fitting  radio luminosity functions}
Several analytical forms have been proposed as good representations of galaxy luminosity functions. The radio luminosity function is often
fitted with a double power-law \citep[e.g.][]{dunlop90,brown01, mauch07} of the form:
\begin{equation}
\Phi(L)={\Phi^{*} \over (L^{*}/L)^{\alpha} + (L^{*}/L)^{\beta}}.\label{eq:power}
\end{equation}
\citet{saunders90} proposed a function which behaves as a power-law at
low radio luminosity and log-normal at high radio luminosities:
\begin{equation}
\Phi(L)=\Phi^{*}({L/L^{*}})^{1-\alpha}\exp(-{1 \over 2\sigma^2}(\log^2(1+L/L^{*}))\label{eq:saunders}
\end{equation}
which has been commonly used to parameterize the radio galaxy luminosity
function \citep[e.g.][]{saunders90,sadler02,smolcic09}. Optical galaxy
luminosity functions are usually represented by a power-law with an
exponential cut-off at the bright-end:
\begin{equation}
\Phi(L)=\Phi^{*}{(L/L^{*})}^{\alpha}\exp(-L/L^{*})\label{eq:schechter}
\end{equation}
 i.e. the Schechter function \citep{schechter76}. This function has the advantage of having one less parameter than the other two.
We tested each of these functional forms in fitting the radio luminosity functions in this paper, including fitting for redshift evolution.
 
In the case of the HERGs all three functional forms perform comparably with, for example, the difference in Akaike Information Criterion \citep[AIC;][]{akaike74}  in every case being
less than 3. An issue which arises in fitting the HERG population is the small number of objects at the bright-end where the number density decreases rapidly.
In the case of the double-power law fit (equation \ref{eq:power}) this manifests itself as a poor constraint on the bright-end slope $\beta$. The sharp drop off in density at high radio luminosities 
and poor statistics means that a good fit can be obtained for arbitrarily large-negative $\beta$ (i.e. as $\beta \rightarrow -\infty$ the power-law slope becomes vertical). Similarly,
for the fitting-function of equation \ref{eq:saunders} the $\sigma$ parameter is poorly constrained. The Schechter function fits the data well at the bright-end, although it is not significantly preferred 
to the other models overall. 

For the LERGs both equation \ref{eq:power} and \ref{eq:saunders} allow good analytical representations of the data, with little difference in the maximum likelihoods. On the
other hand, the \citet{schechter76} function is entirely inappropriate ($\Delta$AIC$\gtrsim 20$ in comparison to the other models when fitting for redshift evolution). For our purposes, the most important
check is that measuring the rate of redshift evolution  (i.e. the $K_{L}$ and $K_{D}$ parameters in equations \ref{eq:luminosity} and \ref{eq:density}) is robust. This is true, since the change in 
the best-fitting value of these parameters between the models is small in comparison to their uncertainties. 

In this paper we have chosen to  use the double power-law representation of equation \ref{eq:power} when fitting the radio luminosity functions. Because of the poor constraints at the bright-end of the HERGs 
we use a logarithmic prior on the $\beta$ parameter. Nevertheless, for the HERGs this parameter essentially only gives an upper limit on the bright-end slope.

\subsection{Fitting the local LERG and HERG populations}
In Figure \ref{fig:rlftype_fit} we show fits of equation \ref{eq:power} to the local LERG and HERG populations.
We add an upper limit point to include the information from a lack of a detection at high radio luminosities.
The value of the upper limit is set such that we can be  68\,per cent  confident the `true' volume density is lower, under the assumption 
of a Poisson distribution \citep{gehrels86}.  The luminosity distributions can be well represented by this function  and we  summarise the best fitting parameter values and
their uncertainty in Table \ref{tab:local_lf_type_fit}.
\begin{figure*}\setcounter{figure}{8}
        \begin{minipage}\textwidth
      \includegraphics[width=8.4cm,angle=0]{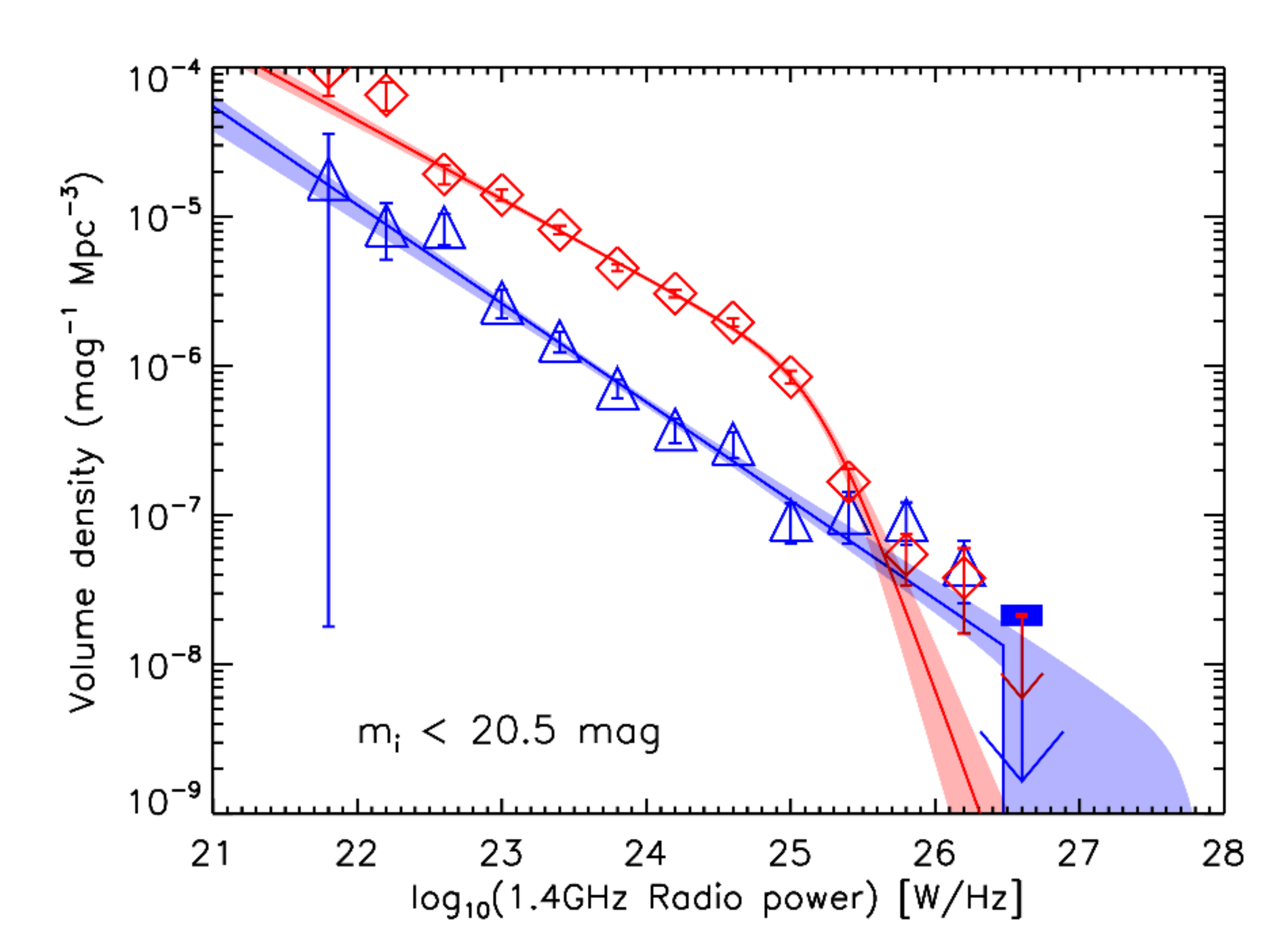}
      \includegraphics[width=8.4cm,angle=0]{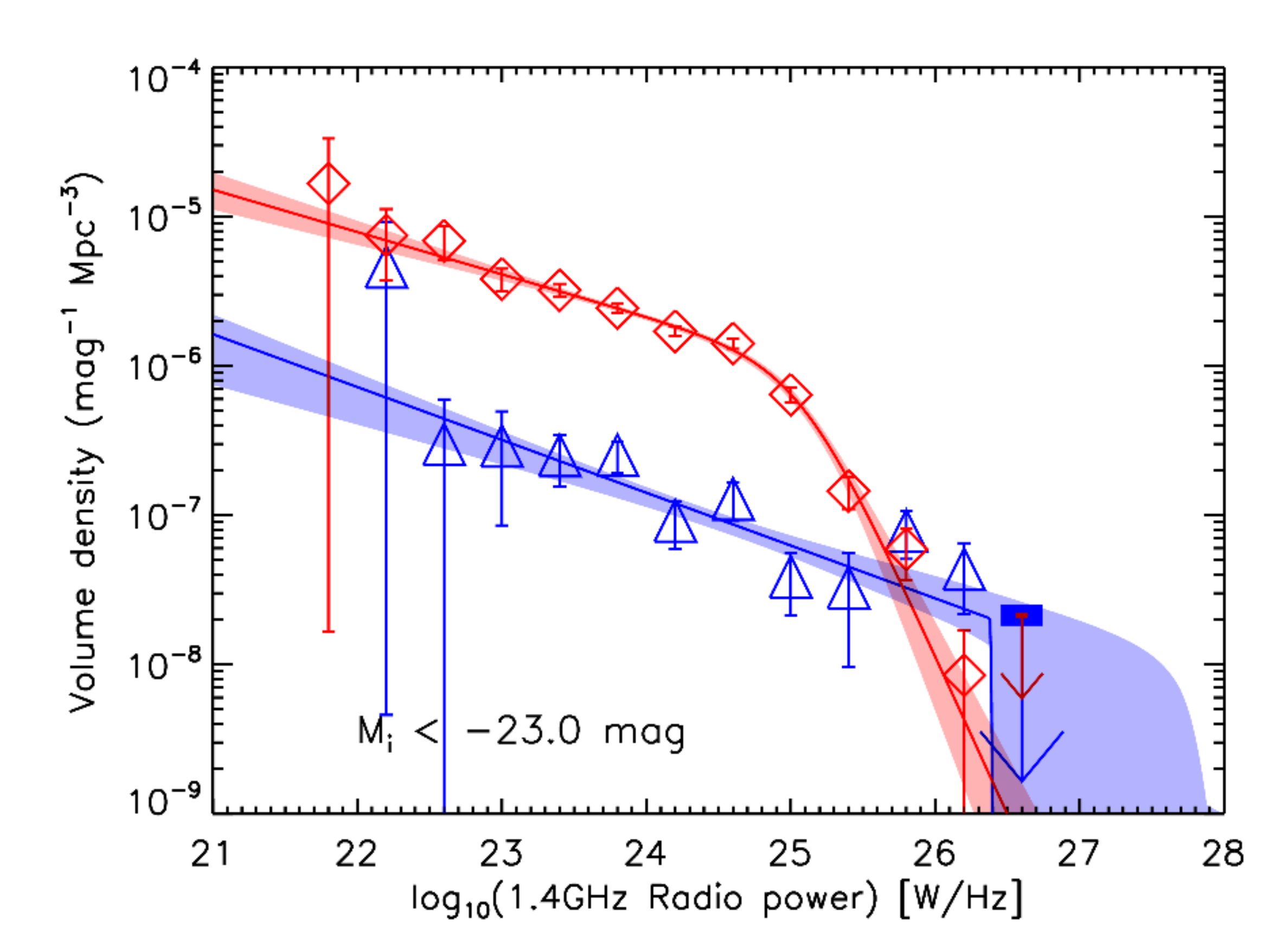}
      \end{minipage}
\caption{\label{fig:rlftype_fit} The radio luminosity function of AGN in
  our $0.005 < z < 0.3$ redshift bin separated into LERGs ({\it red  diamonds and solid line})
  and HERGs ({\it blue  triangles and solid line}). {\it Left panel:}  integrated over $i<20.5$ and
 {\it right panel:}  galaxies with $M_{i}<-23$ only. Over-plotted are
  double power law fits to the data ({\it solid lines}).}
\end{figure*}

\begin{table}
\caption{Best fitted double power-law parameters for the local HERG and LERG luminosity functions.}
\begin{center}
\begin{tabular}{lllll}
\hline
                             &          LERGs        &                         HERGs                      &      LERGs    &              HERGs                         \\
                             &           ($m_{i}<$20.5)      &      ($m_{i}<$20.5)     &         ($M_{i}<$-23) &                ($M_{i}<$-23)                          \\ \hline
$log(C)$                &        $-6.05^{+0.07}_{-0.07}$        &             $-7.87^{+0.19}_{-0.70}$       &        $-5.97^{+0.08}_{-0.08}$         &             $-7.69^{+0.16}_{-0.41}$            \\
$log(P*)$               &        $25.21^{+0.06}_{-0.07}$          &           $26.47^{+1.18}_{-0.23}$      &        $25.08^{+0.08}_{-0.09}$           &             $26.38^{+1.27}_{-0.20}$                  \\
$\alpha$               &        $-0.53^{+0.03}_{-0.07}$          &           $-0.66^{+0.05}_{-0.04} $     &        $-0.28^{+0.05}_{-0.04}$          &              $-0.35^{+0.09}_{-0.06} $                 \\
$\beta$                 &        $-2.67^{+0.42}_{-0.62}$           &                    $ < 0 $                 &       $-0.33^{+0.08}_{-0.07}$           &              $ < 0 $                    \\

\hline
 \end{tabular}
\label{tab:local_lf_type_fit}
\end{center}
\end{table}

\subsection{Redshift evolution}
To examine the redshift evolution of the LERG and HERG population we construct radio luminosity functions for each in multiple redshift
bins. We present the luminosity functions in the same redshift bins as those used for the bivariate luminosity functions shown in Figure
\ref{fig:blf}. That is: $0.005 < z < 0.30$; $0.30 < z  <0.50$; and  $0.5 < z < 0.75$. The luminosity functions for the LERGs and the HERGs are
tabulated in Tables \ref{tab:lerg_lf_z} and \ref{tab:herg_lf_z}. We restrict our analysis to  $M_{i}<-23$; corresponding to the faintest galaxies we can detect at $z=0.75$.
They are plotted in Figure \ref{fig:lerg_herg_redshift}. The LERG luminosity functions, shown
in the  left-hand panels of Figure \ref{fig:lerg_herg_redshift},
exhibit little change in space density with increasing redshift. 
In contrast the HERG luminosity functions ({\it right-hand panels of Figure \ref{fig:lerg_herg_redshift}}) evolve more rapidly. In the highest luminosity bins the evolution presents as a lack of detections
in the lower redshift bins resulting in only upper-limits on the space density of powerful radio galaxies at these redshifts.

The redshift evolution of radio galaxies is often parameterized in terms of pure luminosity evolution, where:

\begin{equation}
 L^{*}(z)=L^{*}(0)(1+z)^{K_{L}}\label{eq:luminosity}
\end{equation}
\citep{boyle88,sadler07}. Substituting into Equation \ref{eq:power} this
results in a fitting function of the form:
\begin{equation}
\Phi(L,z)={C \over ({L^{*}(0)(1+z)^{K_{L}} \over L})^{\alpha} + ({L^{*}(0)(1+z)^{K_{L}} \over L})^{\beta}}.\label{eq:luminosity_fit}
\end{equation}
The other common parameterization is pure density evolution, such that:
\begin{equation}
\Phi (z) = (1+z)^{K_{D}}\Phi (0)\label{eq:density}
\end{equation}
giving:
\begin{equation}
\Phi(L,z)={C (1+z)^{K_{D}}  \over (L^{*}/L)^{\alpha} + (L^{*}/L)^{\beta}}\label{eq:density_fit}
\end{equation}
There is a degeneracy between luminosity and density evolution, especially when the bright population is not tightly constrained \citep{lefloch05,smolcic09}. We therefore do not fit
jointly for luminosity and density evolution but restrict ourselves to the cases above.   

We fitted, using a Markov Chain Monte Carlo (MCMC) method,  both the LERGs ({\it left column of Figure \ref{fig:lerg_herg_redshift}}) and HERGs ({\it right column of Figure \ref{fig:lerg_herg_redshift}}) with  a 
pure density evolution model ({\it top row of Figure \ref{fig:lerg_herg_redshift}}) and a pure luminosity 
evolution model ({\it second row of Figure \ref{fig:lerg_herg_redshift}}).  The evolution in the LERGs can be well represented by either model with the pure luminosity  
evolution marginally preferred with difference in AIC of $\sim 3.5$ (equivalent to a maximum likelihood ratio of $\sim 6$). 
In the case of the HERGs there is a similar preference  for the pure density evolution model with a difference in AIC$\sim 3.1$ equivalent to a maximum likelihood ratio of $\sim 4.5$.
 The best fitting parameters and their uncertainties are summarised in
Table \ref{tab:lf_type_fit}.  Parameterized in this way, the LERGs evolve slowly with redshift as $\sim (1+z)^{0.06^{+0.17}_{-0.18}}$ assuming pure density evolution or $\sim (1+z)^{0.46^{+0.22}_{-0.24}}$ 
assuming pure luminosity evolution.  Under both assumptions this is  consistent with no-evolution within $\sim$2$\sigma$. The HERGs evolve 
 faster than the LERGs with $\sim (1+z)^{2.93^{+0.46}_{-0.47}}$ in the pure density evolution case. If a pure luminosity evolution model is used the parameterised 
redshift dependence is very rapid $\sim  (1+z)^{7.41^{+0.79}_{-1.33}}$.
\begin{figure*}\setcounter{figure}{9}
  \begin{minipage}\textwidth
      \includegraphics[width=8.4cm, angle=0]{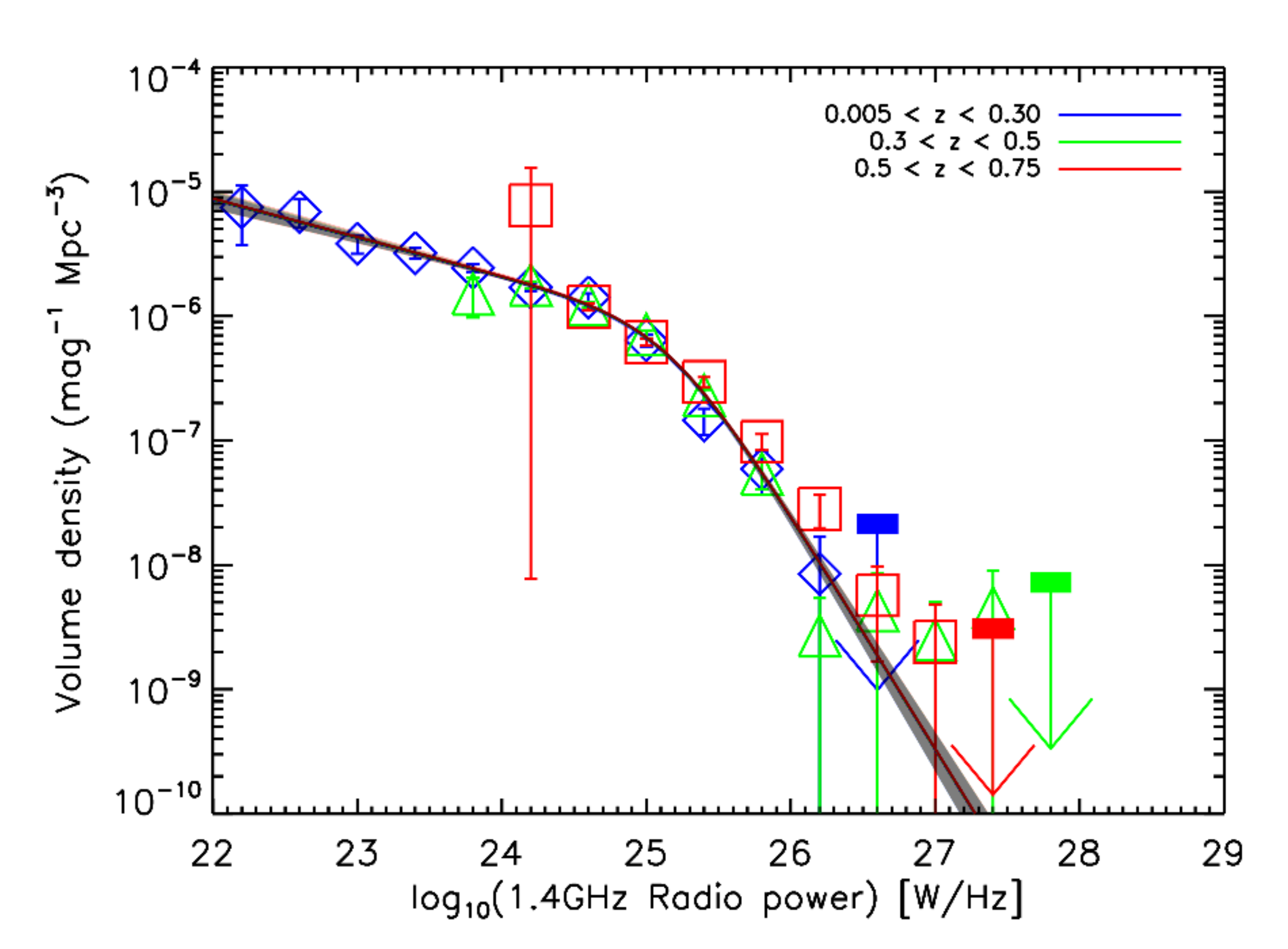}
      \includegraphics[width=8.4cm, angle=0]{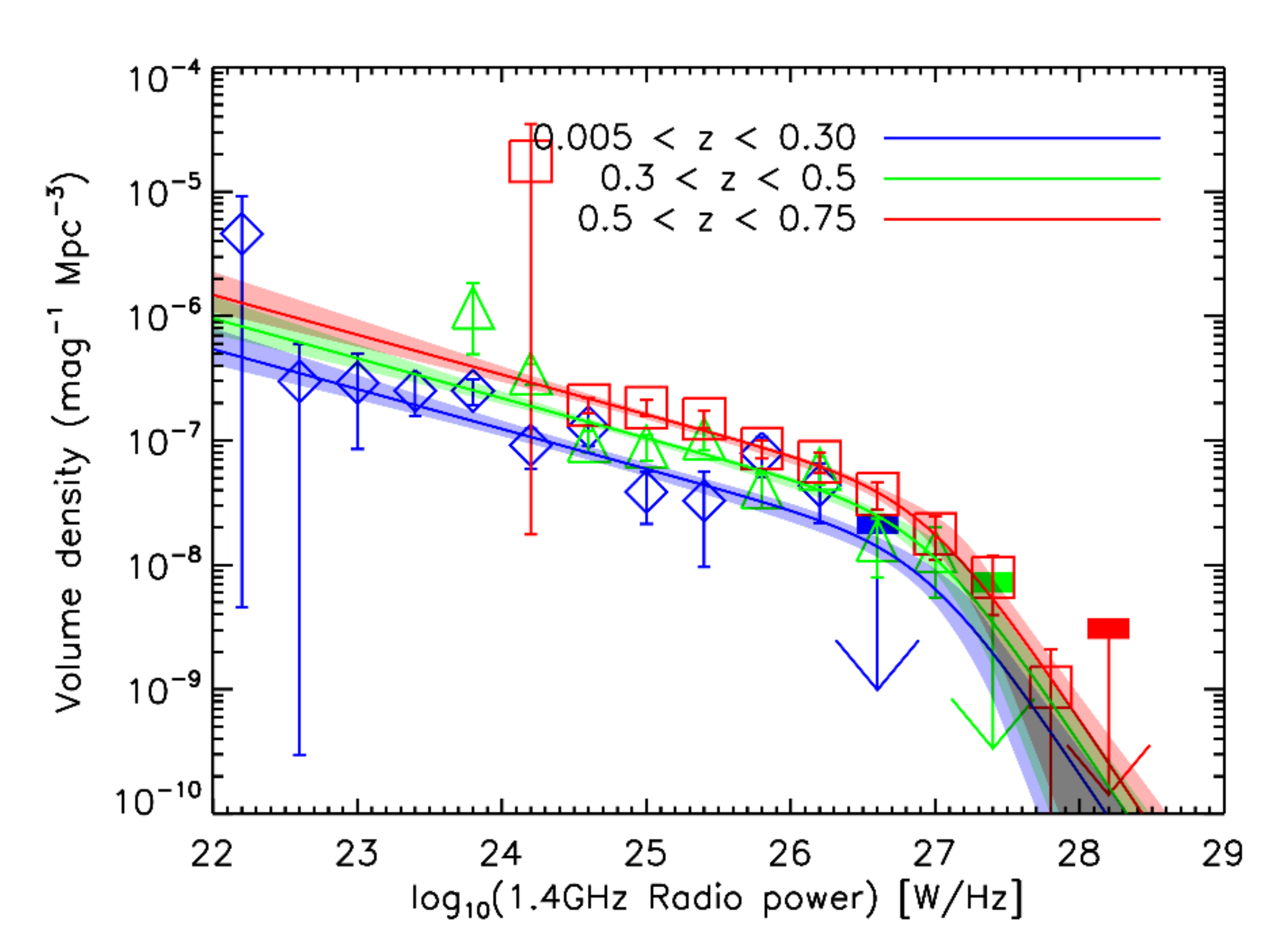}
  \end{minipage}
 \begin{minipage}\textwidth
   \includegraphics[width=8.4cm, angle=0]{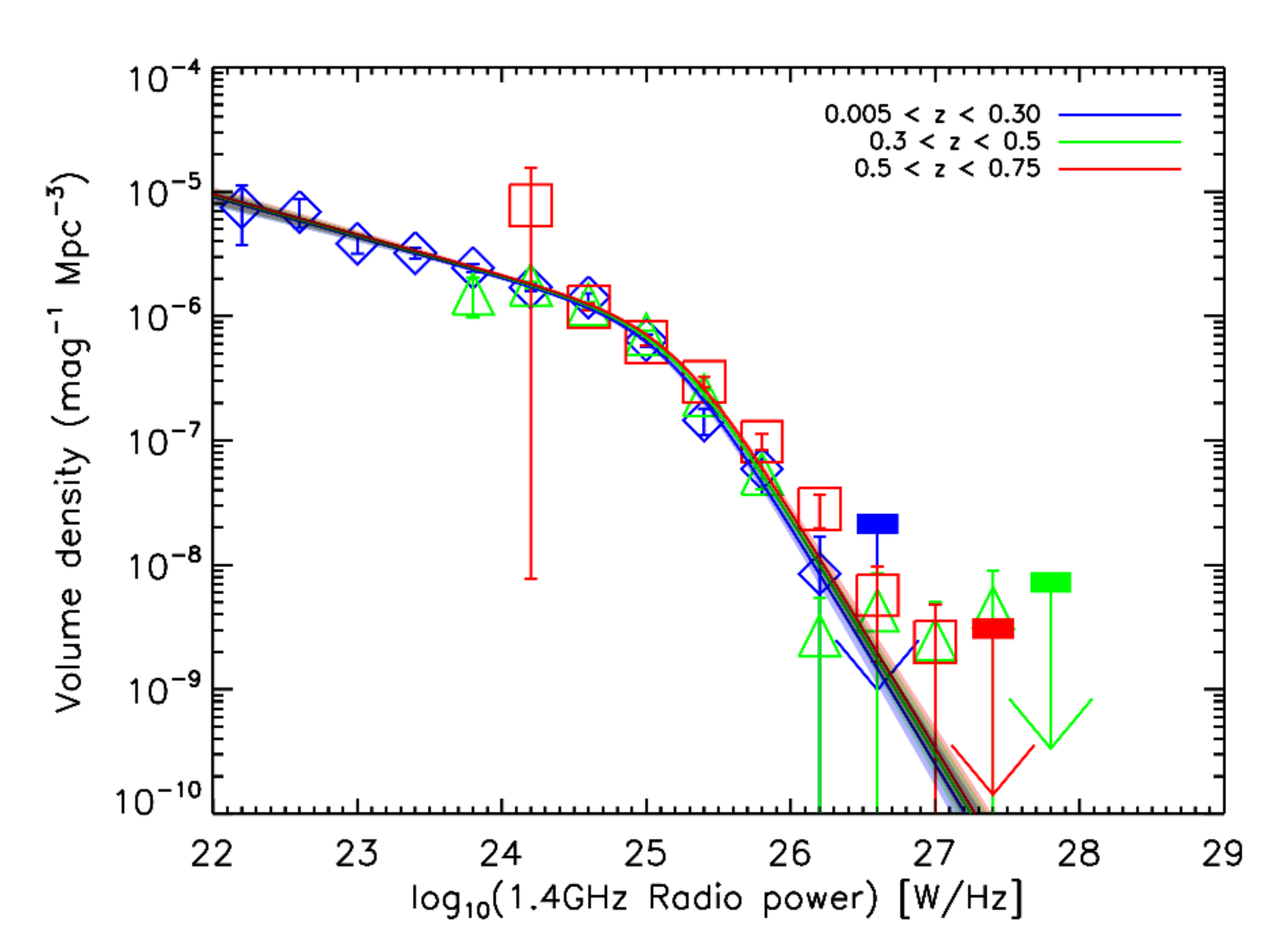}
   \includegraphics[width=8.4cm, angle=0]{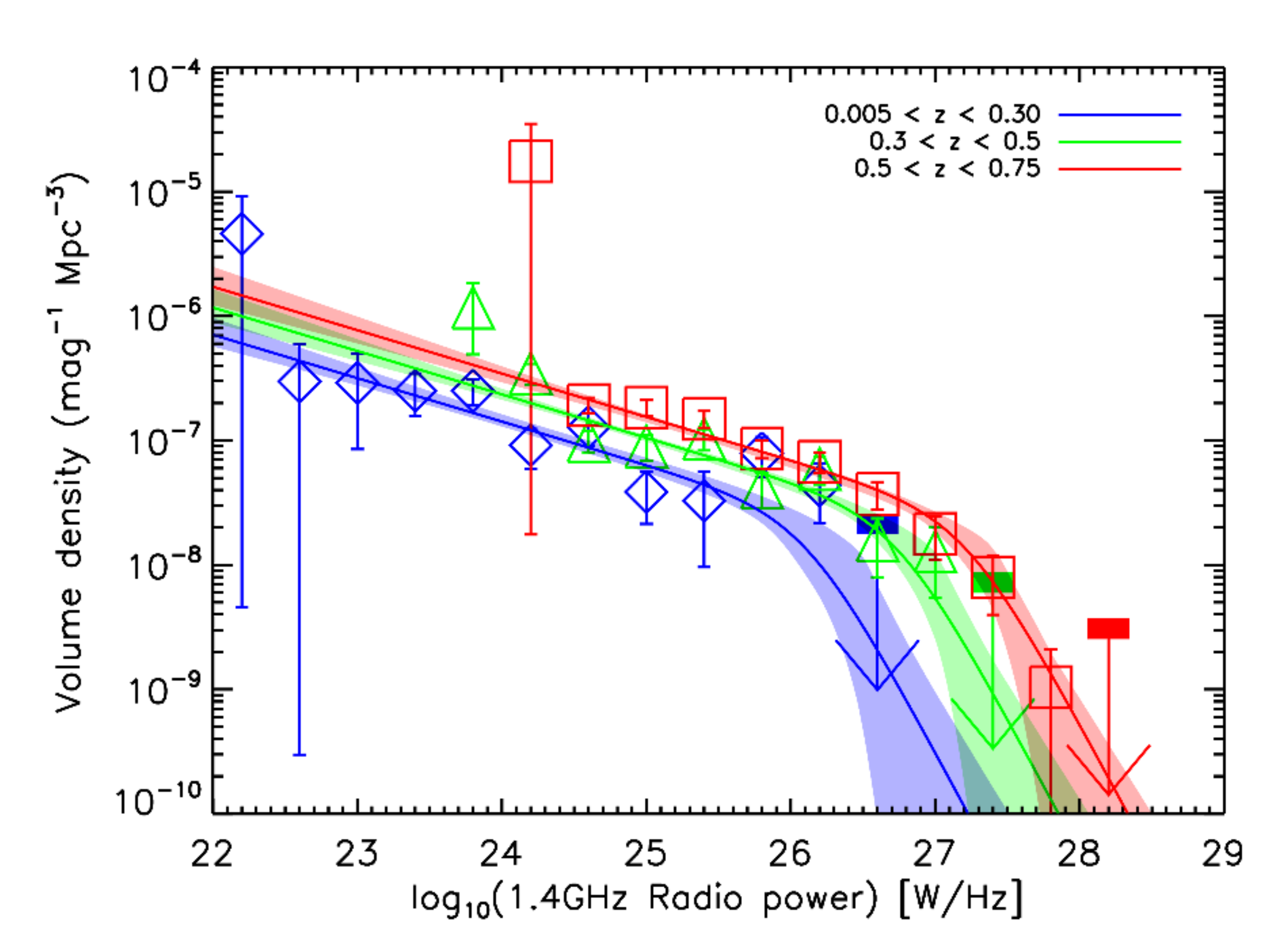}
\end{minipage}
%\begin{minipage}\textwidth
%  \includegraphics[width=6.4cm, angle=90]{diff_lerg}
%  \includegraphics[width=6.4cm, angle=90]{diff_herg}
%\end{minipage}
\caption{\label{fig:lerg_herg_redshift} The radio luminosity function for LERGs ({\it left column}) and HERGs ({\it right column})
separated into three redshift bins: $0.005 < z < 0.30$ ({\it blue}); $0.30 < z  <0.50$ ({\it green}); $0.5 < z < 0.75$ ({\it red}).
The {\it top row} shows the fit to the data ({\it solid lines})  assuming pure density evolution and the {\it bottom row} show the fit to the data
assuming pure luminosity evolution.}
\end{figure*}

\begin{table}
\caption{Best fitted double power-law parameters for the redshift evolution of the LERGs and HERGs.}
\begin{center}

\begin{tabular}{lllll}
\hline
                       &                  &  \hspace{-0.8cm} LERGs          &                        &  \hspace{-0.8cm}     HERGs           \\
                       &      \footnotesize{Density}                   &    \footnotesize{Luminosity}       &    \footnotesize{Density}        &      \footnotesize{Luminosity}                       \\    \hline       
$log(C)$          &     $-6.05^{+0.07}_{-0.06} $          &   $-6.06^{+0.06}_{-0.05}$                          & $-8.04^{+0.17}_{-0.24}$          &  $-7.59^{+0.12}_{-0.19}$      \\
$log(P*)$         &     $25.17^{+0.06}_{-0.06} $            &  $25.12^{+0.06}_{-0.07}$                         & $26.96^{+0.27}_{-0.19}$          &    $25.66^{+0.40}_{-0.23}$      \\    
$\alpha$         &     $-0.31^{+0.04}_{-0.03}$             &   $-0.32^{+0.04}_{-0.03}$                        & $-0.32^{+0.04}_{-0.05}$          &   $0.35^{+0.04}_{-0.05}$      \\
$\beta$           &     $-1.88^{+0.10}_{-0.12}$             &   $-1.92^{+0.12}_{-0.11}$                        & $-1.75^{+0.29}_{-1.40}$           &   $-2.17^{+0.49}_{-4.50}$                \\
$K_{D,L}$            &    $0.06^{+0.17}_{-0.18}$                &  $0.46^{+0.22}_{-0.24}$                      & $2.93^{+0.46}_{-047}$            &  $7.41^{+0.79}_{-1.33}$   \\
\hline
\end{tabular}
\label{tab:lf_type_fit}
\end{center}
\end{table}

\begin{table*}
\caption{Luminosity function of LERGs in 3 redshift bins ($0.005<z<0.30$, $0.30<z<0.50$, $0.50<z<0.75$). Only galaxies with optical I band absolute magnitude M$_{i} <-23$ are included. }
\begin{center}
\begin{tabular}{ccccccc}
\hline\hline  
                               &                 & {\large\hspace{-1.1cm}$0.005<z<0.30$} &          & {\large\hspace{-1.1cm}$0.30<z<0.50$} &          & {\large\hspace{-1.1cm}$0.50<z<0.75$}     \\ 
log$_{10} P_{1.4} $ &    N           &     $\log(\Phi)$                         &    N       &     $\log(\Phi)$                      &    N    &     $\log(\Phi)$                            \\
(W Hz$^{-1}$)      &                  &   (mag$^{-1}$ Mpc$^{-3}$)         &              &  (mag$^{-1}$ Mpc$^{-3}$)       &          &        (mag$^{-1}$ Mpc$^{-3}$)       \\
 21.80                 &    1           &    $-4.78^{+0.30}_{-3.00}$              &              &                                              &           &                                                      \\
 22.20                 &    4          &     $-5.13^{+0.18}_{-0.30}$             &              &                                                &           &                                                     \\
 22.60                 &    15         &      $-5.16^{+0.10}_{-0.13}$             &              &                                                &           &                                                      \\
 23.00                 &     34        &      $-5.42^{+0.07}_{-0.08}$              &              &                                                &           &                                                      \\
 23.40                 &      96       &     $-5.49^{+0.04}_{-0.05}$             &               &                               &           &                                                      \\
 23.80                 &    216           &    $-5.61^{+0.03}_{-0.03}$            &    8           & $-5.82^{+0.13}_{-0.19}$        &         &                   \\                     
24.20                 &     193          &     $-5.77^{+0.03}_{-0.03}$              & 289      &   $-5.75^{+0.02}_{-0.03}$       &     1     &   $-5.11^{+0.30}_{-3.00}$                    \\
 24.60                 &    163              &  $-5.85^{+0.03}_{-0.04}$              & 371     &   $-5.91^{+0.02}_{-0.02}$       &    213       &     $-5.93^{+0.03}_{-0.03}$                \\
 25.00                 &     79           &   $-6.19^{+0.05}_{-0.05}$             &     218     &   $-6.15^{+0.03}_{-0.03}$        &   242    &    $-6.21^{+0.03}_{-0.03}$                 \\
 25.40                 &     17            &    $-6.84^{+0.09}_{-0.12}$             &    77       &  $-6.64^{+0.05}_{-0.05}$       &   121        &      $-6.53^{+0.04}_{-0.04}$               \\
 25.80                 &       7            &   $-7.23^{+0.14}_{-0.21}$              &    17       &  $-7.27^{+0.09}_{-0.12}$     &       45    &     $-7.01^{+0.06}_{-0.07}$                 \\
 26.20                 &       1           & $-8.07^{+0.30}_{-3.00}$                 &      1       &   $-8.57^{+0.30}_{-3.00}$     &       11    &        $-7.55^{+0.11}_{-0.16}$               \\
 26.60                 &                  &                                                       &     1        &  $-8.36^{+0.30}_{-3.00}$      &       2   &      $ -8.24^{+0.23}_{-0.53}$               \\
27.00                 &                  &                                                        &      1       &  $-8.60^{+0.30}_{-3.00}$       &      1     &      $-8.62^{+0.30}_{-2.98}$               \\
27.40                 &                  &                                                        &      1       &  $-8.35^{+0.30}_{-3.00}$       &          &                    \\
\hline
 
\end{tabular}
\end{center}
\label{tab:lerg_lf_z}
\end{table*}

\begin{table*}
\caption{Luminosity function of HERGs in 3 redshift bins ($0.005<z<0.30$, $0.30<z<0.50$, $0.50<z<0.75$). Only galaxies with optical I band absolute magnitude M$_{i}<-23$ are included. }
\begin{center}
\begin{tabular}{ccccccc}
\hline\hline  
                               &                 & {\large\hspace{-1.1cm}$0.005<z<0.30$} &          & {\large\hspace{-1.1cm}$0.30<z<0.50$} &          & {\large\hspace{-1.1cm}$0.50<z<0.75$}     \\ 
log$_{10} P_{1.4} $ &    N           &     $\log(\Phi)$                         &    N       &     $\log(\Phi)$                      &    N    &     $\log(\Phi)$                            \\
(W Hz$^{-1}$)      &                  &   (mag$^{-1}$ Mpc$^{-3}$)         &              &  (mag$^{-1}$ Mpc$^{-3}$)       &          &        (mag$^{-1}$ Mpc$^{-3}$)       \\
 22.20                &     1         &   $-5.34^{+0.30}_{-3.00}$                 &                &                                             &           &                                                              \\
 22.60                &      1        &   $-6.53^{+0.30}_{-3.00}$                  &              &                                               &            &                                                              \\
 23.00                &       2        &   $-6.54^{+0.23}_{-0.53}$                &             &                                                &             &                                                               \\
 23.40                  &      7          &  $-6.60^{+0.14}_{-0.21}$             &           &                                                  &            &                                                               \\
 23.80                  &       18         &   $-6.60^{+0.09}_{-0.12}$           &     3         &      $-5.93^{+0.20}_{-0.37}$            &            &                                                     \\
 24.20                    &      8         &      $-7.04^{+0.13}_{-0.19}$         &     28          &  $-6.46^{+0.08}_{-0.09}$           &       1      &    $-4.76^{+0.30}_{-3.00}$                                                  \\
 24.60                    &      12         &   $-6.89^{+0.11}_{-0.15}$           &     25          &  $-7.00^{+0.08}_{-0.10}$           &     47    &       $-6.72^{+0.06}_{-0.07}$                      \\
 25.00                   &        5          &     $-7.41^{+0.16}_{-0.26}$         &      19         & $-7.05^{+0.09}_{-0.11}$           &     46    &        $-6.74^{+0.06}_{-0.07}$           \\
 25.40                  &         2         &      $-7.48^{+0.23}_{-0.53}$           &     26         &  $-6.99^{+0.08}_{-0.09}$          &     42    &         $-6.83^{+0.06}_{-0.07}$             \\
 25.80                  &         8         &    $-7.10^{+0.13}_{-0.19}$           &       10        &  $-7.38^{+0.12}_{-0.16}$            &    35     &          $-7.06^{+7.06}_{-0.08}$          \\
 26.20                  &         4        &     $-7.36^{+0.18}_{-0.30}$          &        16      &     $-7.23^{+0.10}_{-0.12}$            &    30     &        $-7.17^{+0.07}_{-0.09}$             \\
 26.60                   &                 &                                                  &         4    &        $7.80-^{+0.18}_{-0.30}$         &      16    &      $-7.43^{+0.10}_{-0.12}$                     \\
 27.00                   &                &                                                    &        3        &        $-7.89^{+0.20}_{-0.37}$       &    7       &    $-7.75^{+0.14}_{-0.21} $                    \\
 27.40                   &                &                                                     &                 &                                                &    4         &      $-8.10^{+0.18}_{-0.30}$                                                         \\
27.80                   &                &                                                     &                 &                                                &     1       &      $-8.98^{+0.30}_{-3.02}$                                                         \\
 \hline
 
\end{tabular}
\end{center}
\label{tab:herg_lf_z}
\end{table*}

\section{Discussion}\label{sec:discussion}
It is well established that there is luminosity dependent evolution in the overall radio AGN population, in the sense that the space density of the  high  luminosity population increases more rapidly with 
redshift \citep[e.g.][]{longair66,doroshkevich70,willott01,sadler07,smolcic09}. This differential evolution can be explained by a two population scenario where the LERGs dominate the space density at all but the highest radio luminosity and evolve slowly with redshift,  while the HERGs that dominate at the highest radio luminosities evolve more rapidly \citep{smolcic09,best12,best14}.

The expectation is that the LERGs are hosted by quiescent galaxies and powered by \citet{bondi44} accretion from their hot-gas atmospheres \citep[e.g.][]{hardcastle07,best12}. In this case, a first-order prediction is that the LERGs will evolve in a similar manner to the stellar mass function  of massive quiescent galaxies. In this case a mild decrease with increasing redshift is expected over the redshift range considered here (i.e. $z<0.75$)  evolving down in space density as $\sim (1+z)^{-0.1}$ \citep{best14}.
 In reality the evolution will be complicated by dependence on quantities like the halo hot gas fraction and the cooling function \citep[e.g.][]{croton06} and the average density of the medium into which the radio jets are expanding \citep{best14}. \citet{croton06} predict an almost flat black hole accretion rate density over this redshift range (see Figure \ref{fig:total_power}). In any case, the expectation is we should observe little evolution in our LERG luminosity function out to $z=0.75$, and this is the case with our best-fitting pure density evolution model evolving as $\sim (1+z)^{0.06^{+0.17}_{-0.18}}$ consistent with zero evolution. It should be noted that the evolution measured from our luminosity functions is only for the optically brightest galaxies ($M_{i} < -23.0$), and that our application of an e-correction means we are including the same stellar mass hosts at all redshifts. If the e-correction is not applied our luminosity functions would display a significant positive redshift evolution of the space density of $\sim (1+z)^{0.81^{+0.15}_{-0.16}}$ (assuming pure density evolution).

In this duel accretion mode picture the HERGs are some subset of the optical quasars and Seyfert galaxies. \citet{croom09} measured the evolution of the QSO luminosity function in the interval $0.4<z<2.6$ using the 2SLAQ sample. At the lowest redshifts ($z\lesssim 1$) the number density evolves rapidly in a manner that depends on luminosity --- in the sense that the brightest QSOs evolve faster. At the bright-end ($M_{g} \sim -24$) the space density increases between their lowest redshift bins ($0.40 < z < 0.68$ and $0.68 < z < 1.06$) by a factor of $\sim 3$ which is similar to the change expected from our best fitting pure density model which has space density evolving as $\sim (1+z)^{2.93}$. 

We did not apply an e-correction for the HERGs when constructing the bivariate luminosity functions. Since the HERGs often have ongoing star-formation and it is plausible that the AGN activity and star-formation histories are related ---  it is inappropriate to represent them as having a stellar population which fades with time as it ages. Generally the recent star-formation will contribute much of the optical light, nevertheless there will be some fading of the underlying older stellar population. Applying an e-correction will reduce the magnitude of the evolution in the luminosity function since it shifts higher redshift objects fainter and out of the selection limits. To investigate the maximum difference an e-correction  could make to the evolution of the HERG luminosity function we applied the same e-correction for the HERGs as was applied to the LERGs (which will overestimate the magnitude of the correction) and fitted for the evolution. In the pure density evolution case we find  evolution in the space density of $\sim (1+z)^{2.14^{+0.52}_{-0.63}}$. If a pure luminosity evolution model is used the
redshift dependence is  $\sim  (1+z)^{5.78^{+0.92}_{-2.01}}$.

A clear difference between the luminosity functions of the LERGs and the HERGs is the radio power of the turnover. The bright-end turn down in the HERG luminosity function is $\sim 1$\,dex brighter than that of the LERGs (see e.g. Figure \ref{fig:lerg_herg_redshift} or Table  \ref{tab:lf_type_fit}). 
The bright-end cut-off in the $i$-band optical luminosity function of our HERGs and LERGs is approximately the same (see abscissa values of Figure \ref{fig:blf_type}); implying, to first order, a similar cut-off in the black hole mass function of the two samples. 
The origin of the high power cut off in the radio luminosity function is  likely related to the cut-off in the black-hole mass function modulated in some way by the accretion rate and jet-production efficiency. The Eddington scaled accretion rate in the HERGs  is expected to be 1 to 2\,dex higher than in LERGs \citep{best12,mingo14,turner15,fernandes15}. This difference should be offset to some extent by the higher jet production efficiency in the LERGs \citep[i.e. the ratio of mechanical power to total power;][]{turner15}. In combination the expected mechanical power in units of the Eddington luminosity should be 
$\sim$1\,dex lower for the LERGS than the HERGs \citep{turner15}. This is consistent with  the differences in the characteristic radio power in the HERG and LERG luminosity functions, and is supportive of a picture in which these two populations are powered by different accretion modes.

\subsection{Comparison with Best et al. 2014}
The only previous work on the evolution of the radio luminosity function separated by `accretion mode' using optical spectroscopy was made by \cite{best14} using a composite sample  constructed from eight different radio surveys with a range of flux density limits. This sample contains 211 radio galaxies with redshifts in the range $0.5 <z < 1.0$. As their local comparison sample they use the radio luminosity functions of \cite{best12} but with the HERGs replaced above  $L_{1.4\rm{GHz}}=10^{26}$\,W\,Hz$^{-1}$ with the steep spectrum radio luminosity function from the CoNFIG sample \citep{gendre10,heckman14}.

\subsubsection{Comparison of High Excitation Radio Galaxies}
The  HERG  radio luminosity function of \citet{best14}  exhibits an increase in space density between the local measurement at $z<0.3$
and their high-redshift measurement at $0.5<z<1.0$ of a constant factor of $\sim 7$. This increase is independent of radio luminosity, and well
described by pure density evolution.  Our HERG radio luminosity functions are measured in three redshift bins with $z<0.3$, $0.3<z<0.5$, and
$0.5<z<0.75$. In the best fitting pure density evolution model  the space density  increases as $(1+z)^{2.93^{+0.46}_{-0.47}}$ which, if extrapolated to the  upper redshift 
bin of \citet{best14} corresponds to slightly slower evolution in the space density of a factor of $\sim 4-5$. This is illustrated in Figure \ref{fig:best14_compare_herg} where we compare 
the best-fitting HERG radio luminosity functions of \citet{best14} with our 
HERG radio luminosity function data. This comparison is not quantitatively precise since the classification scheme, and redshift ranges are not identical. The high redshift bin of
\citet{best14} is $0.5 < z < 1.0$ ({\it red  line}) whereas our highest redshift bin is $0.5 < z < 0.75$. Furthermore, we only included the brightest optical galaxies ($M_{i} < -23.0$) when fitting for the redshift 
evolution. If the faint-optical population evolves more rapidly than this would increase the redshift evolution of the space density. 

\begin{figure}\setcounter{figure}{10}
    \includegraphics[width=6.4cm,angle=90]{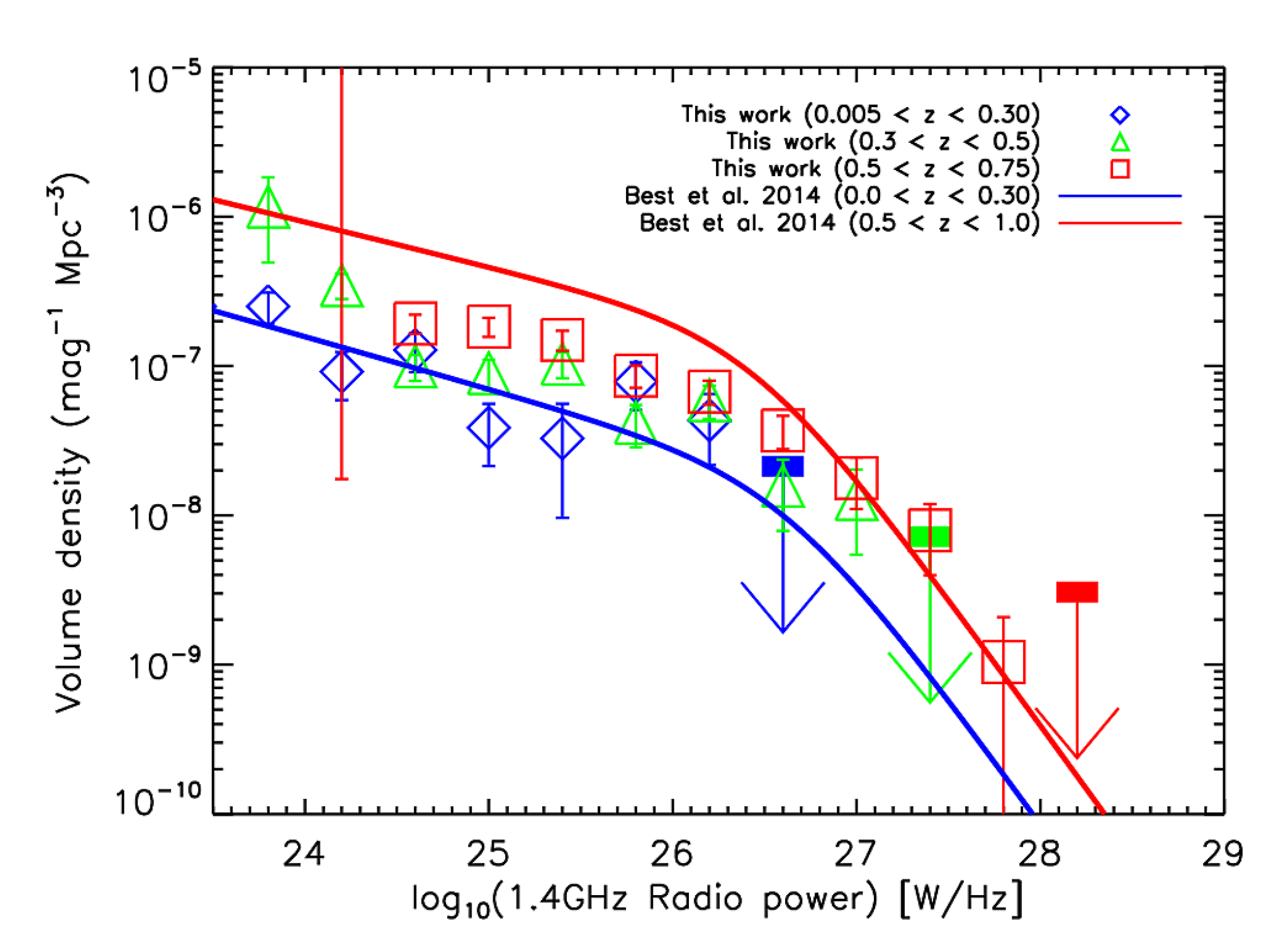}
\caption{\label{fig:best14_compare_herg} A comparison of the radio luminosity functions for the HERGs measured in this work (data points) and the radio luminosity functions of the HERGs from \citet{best14} shown as the solid lines. 
The three redshift intervals for our data are: $0.005 < z < 0.30$ ({\it blue}); $0.30 < z  <0.50$ ({\it green}); $0.5 < z < 0.75$ ({\it red}). Whilst \citet{best14} give their radio luminosity function parameters for $0.005 < z < 0.30$ (blue line) and $0.5 < z < 1.0$ (red  line). }
\end{figure}

\subsubsection{Comparison of Low Excitation Radio Galaxies}
For the  LERGs,  \citet{best14} do find luminosity dependent evolution in the radio luminosity function. At low radio luminosities ($L_{1.4\rm{GHz}}=10^{25}$\,W\,Hz$^{-1}$) they find the space density is nearly constant or increases slowly out to $z\sim$0.5--0.7 and then begins to decrease out to z=1. At high radio luminosity they measure a more rapid increase in space density of LERGs with redshift, rising by factor of $\sim 10$  for the highest radio luminosities. In contrast, we measure much slower evolution in the LERG luminosity function consistent with zero at less than $\sim$2$\sigma$ for both the pure density and pure luminosity evolution models. We do find the luminosity evolution model is a better fit to the data but this preference is marginal. 

These differences, however, are not as significant as they seem since much of the complexity in the  evolution seen by \citet{best14} occurs in their highest redshift bin ($0.7<z<1.0$). In  Figure \ref{fig:best14_compare_lerg} we  compare  our radio luminosity function data points with the best fitting models of \citet{best14}. They are in generally good agreement when comparing the low redshift bins ($z<0.3$) and our high redshift bin ($0.5<z<0.75$) with their intermediate redshift bin ($0.5<z<0.7$). The main difference being \citet{best14} measure a higher space density (a factor of $\sim$2--5) of low luminosity LERGs at both redshifts. 
We do see the beginning of a decrease in space density of the faintest sources with increasing redshift i.e. at faint radio luminosities ($L_{1.4\rm{GHz}} \lesssim
10^{25.5}$\,W\,Hz$^{-1}$) the density measured in our intermediate redshift bin is higher than in the highest redshift
bin. \citet{best14} propose one possible explanation for this decrease as: the space density evolving in line with the density of massive quiescent galaxies and a time delay between the onset of the radio-AGN after the formation of the quiescent host galaxy. 
\begin{figure}\setcounter{figure}{11}
    \includegraphics[width=6.4cm,angle=90]{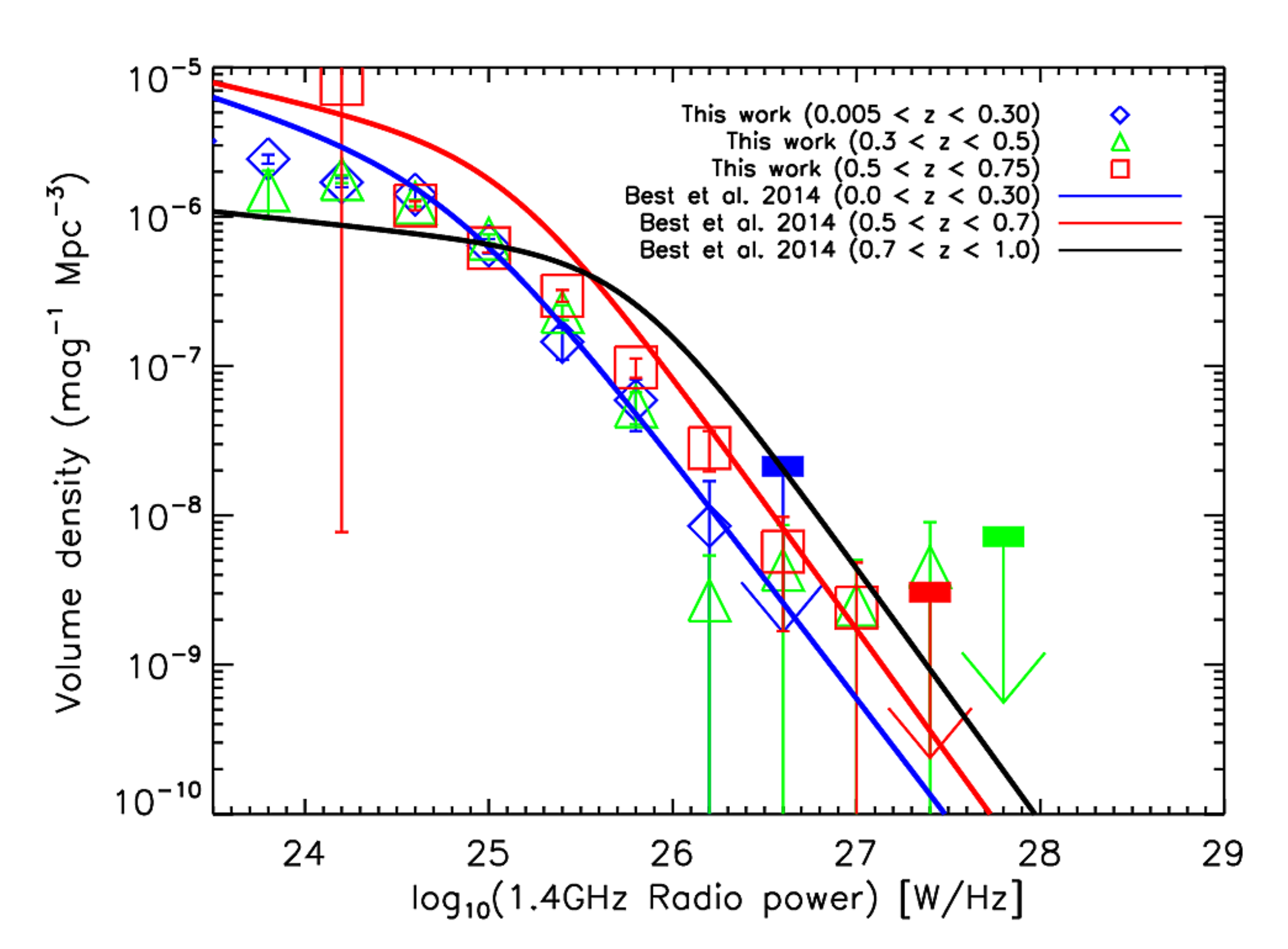}
\caption{\label{fig:best14_compare_lerg}  A comparison of the radio luminosity functions for the LERGs measured in this work (data points) and the radio luminosity functions of the `jet mode'  AGN from \citet{best14} shown as the solid lines. 
The three redshift intervals for our data are: $0.005 < z < 0.30$ ({\it blue}); $0.30 < z  <0.50$ ({\it green}); $0.5 < z < 0.75$ ({\it red}). Whilst \citet{best14} give their radio luminosity function parameters for the LERGs at $0.005 < z < 0.30$ (blue line), $0.5 < z < 0.7$ (red  line) and $0.7<z<1.0$ (black  line). }
\end{figure}

As pointed our earlier, when considering redshift evolution, samples with the same optical brightness constraints should be used otherwise different fractions of the total population will be counted at different redshifts. \citet{best14} do not apply such a constraint instead mitigating such effects by only including surveys with high redshift completeness in their sample. For example,  at a completeness of unity all radio sources are counted and there will be no optical selection effects. \citet{best14} do not state their spectroscopic completeness values for all 8 of the radio surveys used in their combined sample, however, in several cases they restrict their flux density ranges to ensure completeness. We, however, only include the optically brightest galaxies in our sample ($M_{i} < -23$) which has the effect of decreasing the normalisation of the luminosity function. We have already demonstrated that when we do not restrict our sample to $M_{i} < -23$ our local LERG luminosity function agrees with that of \citet[][see Figure \ref{fig:local}]{best14}.

\subsection{Radio mode feedback}
There is substantial evidence that radio jets from LERGs are important in regulating star-formation in massive galaxies and clusters of galaxies by injecting energy into the hot gas atmosphere and inhibiting gas cooling and star formation. This radio-mode AGN feedback can simultaneously  explain the `cooling flow problem', the exponential cut-off in the bright end of the optical galaxy luminosity function and the old stellar populations of the most massive bulges \citep[e.g.][]{croton06,bower06}. Estimates of the energy associated with bubbles and cavities in the intergalactic medium surrounding elliptical galaxies and galaxy clusters can be used to estimate the mechanical power associated with the  radio jets producing the cavities. The empirical correlation of these energies with monochromatic radio luminosity can be used to transform between the two 
\citep{dunn05,rafferty06,birzan08,cavagnolo10}. Although, these relations have large intrinsic scatter of $\sim 0.7$\,dex \citep{cavagnolo10}. Using such relations the monochromatic radio luminosity function can be transformed into a mechanical power density function, and integrated to calculate the total mechanical power (per unit volume) available for radio mode feedback \citep[e.g.][]{best06,smolcic09}. That is, we calculate:
\begin{equation}
  \int \phi (P_{m})\,   P_{\rm m}\,  {\rm d(0.4\log_{10}P_{m})} = {2.5 \over \ln(10)} \int \phi {\rm (P_{m}) \, dP_{ m}}
\end{equation}
where $P_{\rm m}$ is the mechanical power which we calculate from the 1.4\,GHz luminosity using equation (1) of \citet{cavagnolo10}.  The factor of $2.5/ \ln(10)$ comes about since our luminosity function is in units of  ${\rm mag}^{-1}$ rather than units of the natural logarithm.

%UP TO HERE
In Figure \ref{fig:total_power} we show this integral as a function of redshift using our pure density evolution fits to the LERGs and
HERGs. We follow \citet{smolcic09} and integrate above a mechanical power equivalent to L$_{1.4{\rm GHz}}=10^{21}$W\,Hz$^{-1}$.  The
shaded regions illustrate the uncertainties from  the conversion of 1.4\,GHz radio luminosity to mechanical power using the uncertainties
quoted in \citet{cavagnolo10} equation (1) (large shaded regions), and from the uncertainties in our parameter values from fitting the radio
luminosity function; constructed by sampling the posterior
distribution of the parameters obtained from the MCMC fitting (smaller  dark shaded regions). 
\begin{figure*}\setcounter{figure}{12}
        \begin{minipage}\textwidth
      \includegraphics[width=6.4cm, angle=90]{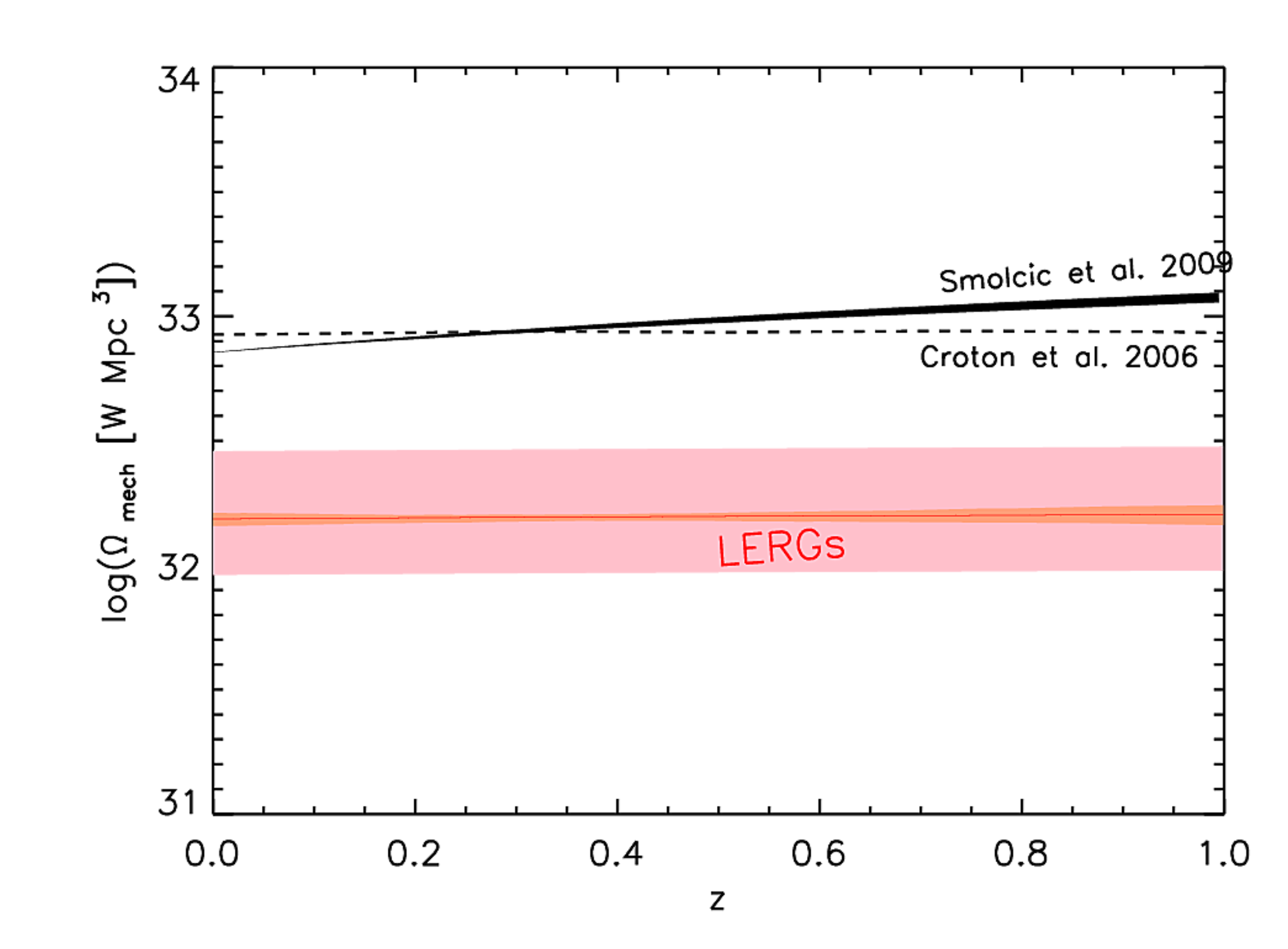}
      \includegraphics[width=6.4cm, angle=90]{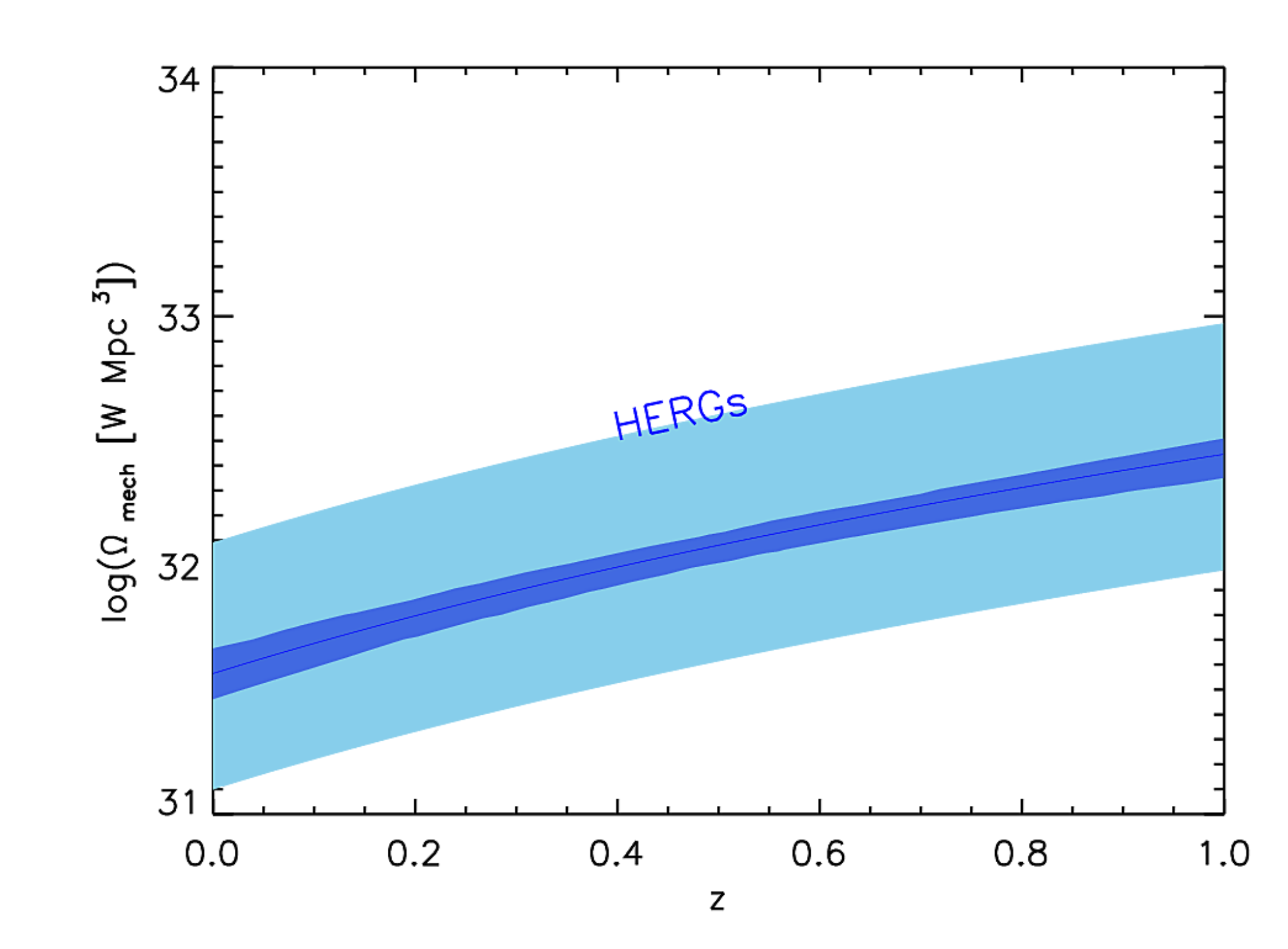}
      \end{minipage}
\caption{\label{fig:total_power} The total mechanical power per unit
  volume as function of redshift estimated from the pure density
  evolution fits to our radio luminosity functions. The conversion to
  mechanical power from 1.4\,GHz luminosity uses the relation of
  \citet{cavagnolo10}. The shaded regions represent the uncertainty
  from the radio luminosity function fits (smaller dark shaded
  regions) and the uncertainty in the  \citet{cavagnolo10} relation
  (the larger shaded regions). The {\it left panel} is for the LERGs
  and the {\it right panel} is for the HERGs. Also over-plotted on the
  LERGs are the prediction from the cosmological model of
  \citet{croton06} and the measurement from
  \citet{smolcic09}.The difference in normalisation between our measurement and that of \citet{smolcic09} can be entirely attributed to our measurement only including the contribution from the brightest optical galaxies (see text for details). }
\end{figure*}

There is little evolution with redshift in the volume density of mechanical power from the LERGs ({\it  left panel} of Figure \ref{fig:total_power}), consistent with the prediction from the cosmological model of \citet{croton06}; shown as the dashed line.   It should be noted these mechanical powers only include emission from massive galaxies since our radio luminosity functions are restricted to $M_{i} < -23$, including fainter optical galaxies will increase the normalisation further. Also, in the left panel of Figure \ref{fig:total_power} we show the mechanical power calculated from fits to the radio luminosity function of low luminosity VLA-COSMOS AGN \citep{smolcic09}. The low luminosity selection means the radio luminosity function should be dominated by LERGs although it will still contain a contribution from the HERGs. The normalisation of our estimate of the total mechanical power is a factor of $\sim 4$ lower than that measured by \citet{smolcic09}. This difference can be attributed to our restriction to only the very brightest optical galaxies ($M_{i} < -23$) causing the normalisation of our luminosity functions to be lower. The difference in normalisation of the luminosity functions accounts for all of  this factor of 4 (c.f. our Figure \ref{fig:lerg_herg_redshift} with \citet{smolcic09} Figure 3).  We also show the evolution of the  mechanical power calculated for the HERGs ({\it right panel}).  The normalisation of this is approximately  an order of magnitude smaller then that of the LERGs at z=0 but becomes comparable at z=1. Since the measurement of \citet{smolcic09} includes both HERGs and LERGs, this evolution in the HERGs can account for the their factor of $\sim 2$ evolution in the total mechanical power out to z=1.

\section{Summary}
We have constructed radio luminosity functions out to a redshift of z=0.75 using a new sample of 5026 radio galaxies with 1.4\,GHz flux density: $S_{1.4{\rm GHz} }> 2.8$\, mJy and optical magnitude: $m_{i}<20.5$\, mag. These radio galaxies have confirmed spectroscopic redshifts. The optical spectra are also used to classify the radio galaxies as star-forming or AGN. The AGN are further subdivided into High Excitation Radio Galaxies (HERGS) or Low Excitation Radio Galaxies (LERGS). Using a subset of the brightest optical galaxies with $M_{i}<-23$ we characterise the evolution in the radio luminosity function of these objects. We find:
\begin{itemize}
\item The space density of the LERGs exhibits little evolution  with redshift. The LERGs evolve as $\sim (1+z)^{0.06^{+0.17}_{-0.18}}$ under the assumption of pure density evolution or  $\sim (1+z)^{0.46^{+0.22}_{-0.24}}$ under the assumption of pure luminosity evolution. Both are consistent with zero evolution at less than 2$\sigma$. Since the LERGs dominate the number density of radio AGN at low radio luminosity this result is consistent with the mild evolution seen in the total radio luminosity function at low radio power. 

\item The HERGs evolve more rapidly  best fitted by $\sim (1+z)^{2.93^{+0.46}_{-0.47}}$ assuming pure density evolution or $\sim (1+z)^{7.41^{+0.79}_{-1.33}}$ under the pure luminosity evolution assumption.  Since the HERGs dominate the space density at only the highest luminosities, this is consistent with the more rapid evolution of bright radio galaxies observed in the overall radio AGN population.

\item The bright-end turn down 
in the radio luminosity function occurs at a significantly higher power ($\gtrsim 1$\,dex) for the HERG population than the LERG population. This is 
consistent with the two populations representing fundamentally different accretion modes. 

\item Converting the LERG luminosity function to a mechanical power density function using empirical relations and integrating, results in a total mechanical power  per unit volume available for radio mode feedback that remains roughly constant out  to z=0.75. This is consistent with cosmological models.

\end{itemize}

\section*{Acknowledgements}
We are grateful to the anonymous referee for insightful comments which greatly improved this paper. 

EMS and SMC acknowledge the financial support of the Australian
Research Council through Discovery Project grants DP1093086 and DP
130103198. SMC acknowledges the support of an Australian Research
Council Future Fellowship (FT100100457). S.B. acknowledges the funding support from the Australian Research
Council through a Future Fellowship (FT140101166). 

Funding for the Sloan Digital Sky Survey (SDSS) and SDSS-II has been provided by the Alfred P. Sloan Foundation, the Participating Institutions, the National Science Foundation, the U.S. Department of Energy, the National Aeronautics and Space Administration, the Japanese Monbukagakusho, and the Max Planck Society, and the Higher Education Funding Council for England. The SDSS Web site is http://www.sdss.org/.

The SDSS is managed by the Astrophysical Research Consortium (ARC) for the Participating Institutions. The Participating Institutions are the American Museum of Natural History, Astrophysical Institute Potsdam, University of Basel, University of Cambridge, Case Western Reserve University, The University of Chicago, Drexel University, Fermilab, the Institute for Advanced Study, the Japan Participation Group, The Johns Hopkins University, the Joint Institute for Nuclear Astrophysics, the Kavli Institute for Particle Astrophysics and Cosmology, the Korean Scientist Group, the Chinese Academy of Sciences (LAMOST), Los Alamos National Laboratory, the Max-Planck-Institute for Astronomy (MPIA), the Max-Planck-Institute for Astrophysics (MPA), New Mexico State University, Ohio State University, University of Pittsburgh, University of Portsmouth, Princeton University, the United States Naval Observatory, and the University of Washington.

The WiggleZ project wishes to acknowledge financial support from The Australian Research Council (grants DP0772084 and LX0881951 directly
for the WiggleZ project, and grant LE0668442 for programming support).

GALEX is a NASA Small Explorer, launched in 2003 April. We gratefully acknowledge NASA's support for construction, operation and science analysis for the GALEX mission, developed in cooperation
with the Centre National d’Etudes Spatiales of France and the Korean Ministry of Science and Technology

GAMA is a joint European-Australian project based around a spectroscopic campaign using the Anglo-Australian Telescope. The GAMA input catalogue is based on data taken from the Sloan Digital Sky Survey and the UKIRT Infrared Deep Sky Survey. Complementary imaging of the GAMA regions is being obtained by a number of independent survey programmes including GALEX MIS, VST KIDS, VISTA VIKING, WISE, Herschel-ATLAS, GMRT and ASKAP providing UV to radio coverage. GAMA is funded by the STFC (UK), the ARC (Australia), the AAO, and the participating institutions. The GAMA website is http://www.gama-survey.org/.

%%%%%%%%%%%%%%%%%%%%%%%%%%%%%%%%%%%%%%%%%%%%%%%%%%



\bibliographystyle{mnras}
\bibliography{/Users/mpracy/work/atca/single_file/notes/references}

% Don't change these lines
\bsp	% typesetting comment
\label{lastpage}
\end{document}